\documentclass[preprint2]{aastex}
\newcommand{\etal}{{et \it al.}}

\newcommand{\lsun}{L$_{\odot}$}

\slugcomment{DRAFT v2000.11.01}					

\begin{document}

\title{HST-NICMOS Observations of M31's Metal Rich Globular Clusters
and Their Surrounding Fields\altaffilmark{1} I. Techniques}

\author{Andrew W. Stephens \& Jay A. Frogel}
\affil{The Ohio State University, Department of Astronomy}
\affil{140 West 18th Avenue, Columbus, OH  43210}

\author{Wendy Freedman \& Carme Gallart}
\affil{Carnegie Observatories}

\author{Pascale Jablonka}
\affil{Observatoire de Paris-Meudon}

\author{Sergio Ortolani}
\affil{Universit\`a di Padova}

\author{Alvio Renzini}
\affil{European Southern Observatory}

\author{R. Michael Rich}
\affil{University of California at Los Angeles}

\and

\author{Roger Davies}
\affil{University of Durham}

\altaffiltext{1}{Based on observations with the NASA/ESA Hubble Space Telescope 
obtained at the Space Telescope Science Institute, 
which is operated by AURA for NASA under contract NAS5-26555.}

\begin{abstract}

Astronomers are always anxious to push their observations to the limit
-- basing results on objects at the detection threshold, spectral
features barely stronger than the noise, or photometry in very crowded
regions.  In this paper we present a careful analysis of photometry in
crowded regions, and show how image blending affects the results and
interpretation of such data.  Although this analysis is specifically for
our NICMOS observations in M31, the techniques we develop can be applied
to any imaging data taken in crowded fields; we show how the effects of
image blending will even limit NGST.

We have obtained HST-NICMOS observations of five of M31's most metal
rich globular clusters.  These data allow photometry of individual stars
in the clusters and their surrounding fields.  However, to achieve our
goals -- obtain accurate luminosity functions to compare with their
Galactic counterparts, determine metallicities from the slope of the
giant branch, identify long period variables, and estimate ages from the
AGB tip luminosity, we must be able to disentangle the true properties
of the population from the observational effects associated with
measurements made in very crowded fields.

We thus use three different techniques to analyze the effects of
crowding on our data, including the insertion of artificial stars
(traditional completeness tests) and the creation of completely
artificial clusters.  These computer simulations have proven invaluable
in interpreting our data.  They are used to derive threshold- and
critical-blending radii for each cluster, which determine how close to
the cluster center reliable photometry can be achieved.  The simulations
also allow us to quantify and correct for the effects of blending on the
slope and width of the RGB at different surface brightness levels.  We
then use these results to estimate the limits blending will place on
future space-based observations.

\end{abstract}


%
%
\section{Introduction} \label{sec:introduction}

The main objective of our observations was to obtain physical parameters
for a selection of metal-rich globular clusters in M31.  These
parameters are usually straight-forward to derive from color-magnitude
diagrams (CMDs), however, we have found the effects of crowding to be
particularly severe.  We therefore present our analysis in two parts.
This paper describes in detail the effects of crowding, how we quantify
them, the techniques used to correct for them, and their implications
for future space-based observations such as with NGST.  A second paper
\cite[][hereafter Paper II]{SFFJ2001} presents the science obtained from
our observations using these techniques.

The general effects of crowding are well known, although not always
accounted for.  These effects include reduced photometric accuracy which
broadens the CMD and smears the luminosity function (LF), reduced
positional accuracy \citep{Hog2000}, artificial brightening and shifts
in the measured colors \citep{AG1995}, and in severe cases, the creation
of objects brighter than anything in the parent population through
random groupings of many stars \citep{Ren1998, DTFA1993}.  If these
effects are not taken into account, they can result in a mistaken spread
in metallicity assumed from the RGB width, an incorrect mass function
extrapolated from the LF, or a false age estimation from the AGB peak
luminosity.

This paper is organized as follows.  Section \ref{sec:observations}
presents the details of our observations, and section
\ref{sec:data_reduction} explains the reduction procedures.  Section
\ref{sec:artificial_star_tests} describes the different simulations in
detail.  Section \ref{sec:results} presents the results which are
applied to the cluster data in Paper II, and section
\ref{sec:implications} shows how these results can be applied to future
space-based observations, such as the NGST.  We summarize our main
conclusions in section \ref{sec:conclusions}.

%
%
\section{Observations} \label{sec:observations}

We have obtained HST NICMOS images of five of M31's metal rich globular
clusters and their surrounding fields (Cycle 7; Program ID 7826).  The
clusters are G1, G170, G174, G177 \& G280; however most of the
discussion in this paper will focus on G280 which has the densest core,
one of the sparsest surrounding fields, and is dispersed enough to allow
accurate cluster star measurements in its outer regions.

Our observations were taken with the NICMOS camera 2 (NIC2) which has a
plate scale of $\sim$ 0\farcs0757 pixel$^{-1}$ and a field of view of
19\farcs4 on a side (376 arcsec$^2$).  The NICMOS focus was set at the
compromise position 1-2, which optimizes the focus for simultaneous
observations with cameras 1 and 2.  All of our observations used the
{\sc multiaccum} mode \citep{Mac1997} because of its optimization of the
detector's dynamic range and cosmic ray rejection.

Each of our targets was observed through three filters: F110W (0.8--1.4
\micron), F160W (1.4--1.8 \micron), and F222M (2.15--2.30 \micron).
These filters are close to the standard ground-based $J$, $H$, \& $K$
filters.  Each globular cluster was observed over three orbits of HST,
with $\sim 42$ minutes of observations per orbit.  This yielded total
integration times of 1920s in F110W, 3328s in F160W, and 2304s in F222M
(see Table \ref{tab:filters}).

We implemented a spiral dither pattern with 4 positions to compensate
for imperfections in the infrared array.  The dither steps were 0\farcs4
for the $J$ and $K$ band images, and 5\farcs0 for the $H$ band images.
Thus the full size of the combined dithered images are $\sim 20''$ in
$J$ and $K$, and $\sim 24''$ in $H$.  We used the predefined sample
sequences {\sc step32} with 22 samples in $J$ and 25 samples in $K$, and
{\sc step64} with 21 samples in $H$ \cite{Mac1997}.

\begin{figure}
\epsscale{1}
\plotone{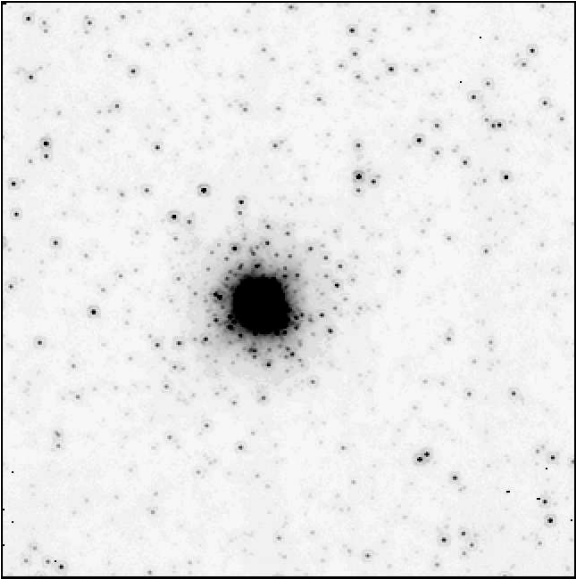}
\figcaption{G280 -- F160W ($H$-band) combination of all 4 dithers; 3328s total exposure.
\label{fig:g280}}
\end{figure}

The $H$-band image of G280 is shown in Figure \ref{fig:g280}.  This is
the combination of 4 dither positions, and is $\sim 24''$ on a side.

\begin{figure}
\figurenum{1b}
\epsscale{1}
\plotone{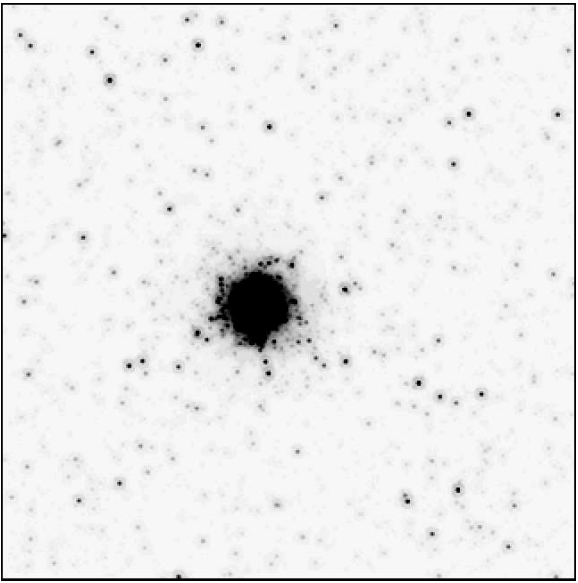}
\figcaption{A280 -- The artificial analog of the G280 frame.
\label{fig:a280}}
\end{figure}

%
%
\section{Data Reduction} \label{sec:data_reduction}

Our data were reduced with the STScI pipeline supplemented by the IRAF
NICPROTO package (May 1999) to eliminate any residual bias (the
``pedestal'' effect).  To account for any bias shift which may have
occurred in the NICMOS reduction (arbitrary amounts get added or
subtracted during pedestal removal), or unsubtracted background
contribution, we scale our surface brightness by adding a constant to
the entire frame to agree with the surface brightness from
\citet{Ken1987}.  As an example, the procedure for the G280 frame was as
follows.  Taking the observed position of G280 as $20.5'$ from the
center of M31, and 34\degr\ from the major axis, we interpolate Kent's
data to find an equivalent major axis distance of $\sim 23.8'$ from
M31's center.  At this distance the total $r$-band surface brightness is
$\mu_r = 21.2$ magnitudes arcsecond$^{-2}$.  Assuming an $(r-K)$ color
of 2.9 from \citet{TDFD1994} for the disk of a galaxy similar to M31, we
estimate the $K$-band surface brightness to be 18.3 magnitudes
arcsecond$^{-2}$.  We then added a constant 0.045 to the count-rate in
each pixel of our frame to bring the average surface brightness of the
frame up to this level (measured $> \sim 8''$ around the cluster).

Object detection was performed with DAOFIND on a combined image made up
of all the dithers of all the bands (12 images in total).  PSFs were
constructed with DAOPHOT for each of the four dithers, then averaged
together to create a single PSF for each band of each target (the
average FWHM of each band is listed in Table \ref{tab:filters}).
Instrumental magnitudes were measured using the ALLFRAME PSF fitting
software package \citep{Ste1994}, which simultaneously fits PSFs to all
stars on all dithers.  DAOGROW \cite{Ste1990} was used to determine the
best magnitude in a $0.5''$ radius aperture, which we then converted to
the CIT/CTIO system using the transformation equations of
\citet{SFOD2000}.

\begin{deluxetable}{cccc}
\tablewidth{5.5cm}
\tablecaption{NICMOS Filters}
\tabletypesize{\footnotesize}
\tablehead{
\colhead{Filter}	&
\colhead{Exposure} 	&
\multicolumn{2}{c}{FWHM} \\
\colhead{}		&
\colhead{(s)}		&
\colhead{(pix)}		&
\colhead{($''$)}	}
\startdata
F110W	& 1920	& 1.65	& 0.13	\\
F160W	& 3328	& 1.95	& 0.15	\\
F222M	& 2304	& 2.45	& 0.19	\\
\enddata
\label{tab:filters}
\end{deluxetable}

%
%
\section{Artificial Star Tests} \label{sec:artificial_star_tests}

The measurements of stars in the M31 clusters and their surrounding
fields will be affected by crowding due to the high stellar density.  We
have performed three types of tests to evaluate the effects of crowding
on our photometry and to quantify at what surface brightness levels we
can achieve accurate measurements.  We compare these techniques with one
another (\S \ref{sec:comparison}), and with predictions from first
principles (\S \ref{sec:simulations_vs_theory}).

%
%
\subsection{Traditional Completeness Tests} \label{sec:traditional_completeness_tests}

Our first attempt to determine the effects of crowding, as well as to
quantify our photometric completeness, used traditional artificial star
tests.  We added stars of a fixed magnitude to each individual dither
using ADDSTAR \citep{Ste1987}, re-processed the frames, and recorded the
recovered magnitudes.  Since we are interested in not only the
completeness, but also the effects of blending, we did not incorporate a
delta magnitude criterion for the recovered stars.

We repeated this procedure for nine different input magnitudes
$(-1<M_H<-9)$.  Each trial added only 81 stars per frame to minimize
additional crowding.  These 81 stars were arranged in a grid with
spacing of $2 \times R_{PSF} + 2 =18$ pixels to avoid self-crowding.
The grid of input stars was then shifted by 3 pixels and the process
repeated so that the entire cluster was covered in 36 trials for each of
the 9 input magnitudes (in the end an artificial star of each magnitude
had been added to every 9th pixel in the region of the cluster).  We
thus input 2916 artificial stars per magnitude.

\begin{figure}
\epsscale{1}
\plotone{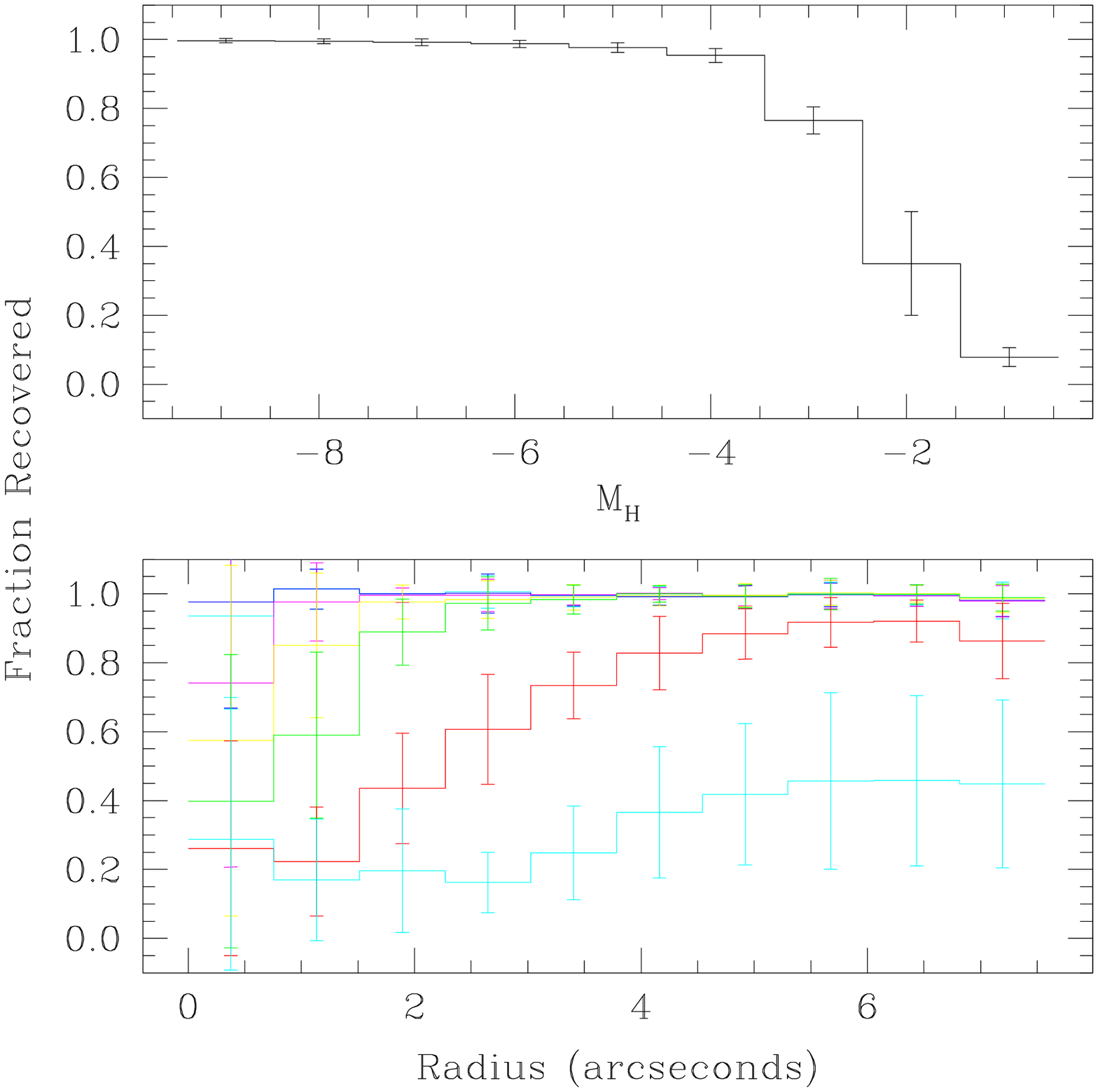}
\figcaption{
G280 fractional completeness -- (a) as a function of input magnitude for
the whole frame, and (b) as a function of distance from the center of
the cluster for each input magnitude.  Each histogram shows the result
of a different input magnitude ranging from $M_H=-8$ (top histogram) to
$M_H=-2$ (bottom histogram)
\label{fig:completeness}}
\end{figure}

Not surprisingly, Figure \ref{fig:completeness} demonstrates that the
completeness is a function of both input magnitude and distance from the
cluster center.  The top panel shows the fractional completeness for the
entire G280 frame as a function of input magnitude.  Here each bin
represents a total of 2916 artificial stars added 81 stars at a time.
The error bars represent the dispersion in the number of recovered stars
over the 36 trials.  The bottom panel shows the completeness of each
magnitude trial as a function of position from the center of the
cluster.  Here the faintest input magnitude plotted is $M_H = -2$, which
is only $\sim 40$\% complete in the outer regions, and drops to a value
consistent with zero at $\sim 2''$.  Input magnitudes from $M_H=-4$ to
$-8$ inclusive, show essentially 100\% completeness for radii $\geq 2.5''$.  
Input magnitudes $M_H=-7$ and $-8$ remain 100\% complete down
to $0''$ radius

\begin{figure}
\epsscale{1}
\plotone{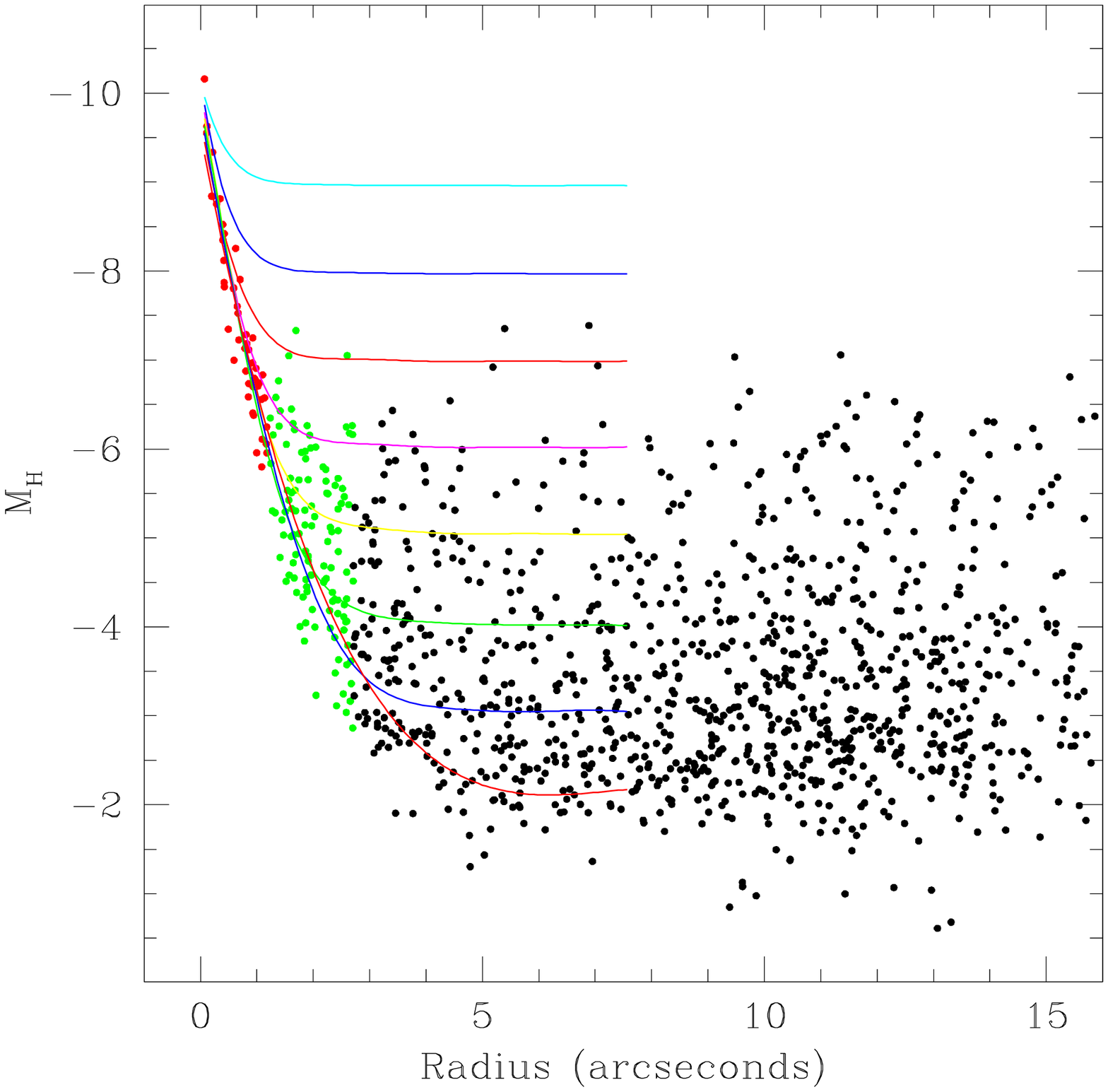}
\figcaption{
Stars measured on the observed G280 frame, and the loci of the {\it
recovered} artificial stars for each input magnitude from the
traditional completeness tests.  There is good correspondence between
the input and recovered magnitudes at large radii, so the input
magnitude may be inferred from the right hand side of each line.  The
top line is the locus of measurements resulting from stars input with
$M_H = -9$, and the bottom line is from $M_H = -2$.  The upward curving
of these lines results from the increasing difference between input and
recovered magnitudes at decreasing radii.
\label{fig:artstarfits}}
\end{figure}

Figure \ref{fig:artstarfits} illustrates an important aspect of the
crowding problem not shown by Figure \ref{fig:completeness}; the
difference between input and recovered magnitudes.  As will be seen, for
our purposes, the traditional completeness plots (Fig
\ref{fig:completeness}) are largely irrelevant.  We are mainly concerned
with accurately measuring the upper red giant branch (RGB), and thus
blending at the bright end, and not completeness at the faint end, is
our primary worry.  Figure \ref{fig:artstarfits} shows the effects of
blending as a function of radial distance from the cluster.  The dots
are the measured objects in the G280 frame.  The lines are the locus of
the recovered artificial star magnitudes.  At large radii $(\sim 5'')$
there is good correspondence between the input and the recovered
(plotted) magnitudes (thus the actual value of each set of input
magnitudes can be obtained from the level of each line at $r>5''$).
However at small radii, what is measured can be several magnitudes above
what was input.  Regardless of the star's input brightness, if a star is
recovered near the center of the cluster, it is recovered with the
luminosity of the object on which it fell (if it does not fall near a
bright star, then it is most likely not recovered).

Note that the brightest objects measured on the real frame are all
within the central few arcseconds, and that the upward curving loci of
the artificial stars closely mimic the distribution of those bright
stars measured on the real frame.  This is indicative of the blended
nature of these objects, but does not reveal their true luminosities, as
{\em all} the loci of the artificial stars pass through the brightest
measured objects at the center of the cluster.

%
%
\subsection{Artificial Clusters} \label{artificial_clusters}

\subsubsection{Generation of the Artificial clusters \& Fields}

From the photometric results in Figure \ref{fig:artstarfits}, we see
that we measure predominantly very bright objects in crowded regions.
The artificial star tests showed that the measurement of faint stars
injected into these crowded regions will be significantly altered by the
severe crowding, and these faint stars will usually be absorbed into the
nearest bright object.  To further investigate the nature of these
bright objects, we decided that more tests were necessary.  However,
heeding the warning of \citet{DTFA1993} that typical artificial star
experiments cannot reproduce the true effects of severe image crowding,
and following the lead of \citet{RMG1993}, we initiated our own set of
``non--standard'' artificial star experiments.

These experiments involved the simulation of the entire cluster and
surrounding field using a reasonable estimate of the expected cluster
star properties.  Starting with a blank frame with the appropriate noise
characteristics, we randomly added cluster and field stars, using the
same PSFs determined from the G280 cluster, with any negative values set
to zero.  Both the cluster and field stars follow the luminosity
function and colors observed in the Galactic Bulge, as we expect the
cluster stars to have similar properties, and the bulge LF is complete
to faint levels in the IR.  The LF is a power law with a slope of 0.278,
extending from $-7.2<M_K<5.8$ ($7.0 < K < 20$ at Baade's Window)
\citep{TFT1995, DTFA1993}.  Later we will examine the effect of changes
in the assumed input LF.  Colors are derived from a combination of
observations of the RGB slope for $M_K<2.8$ \citep{TFT1995}, and from
stellar models for $M_K > 2.8$ (Cassisi).  Cluster stars were added
according to the radial surface brightness profile measured in the real
cluster until the integrated magnitude of the cluster matched that of
the cluster being modeled.  Note that the measurement surface brightness
is independent of crowding, and as long as there are not saturated
pixels, which there were not, the measured SB is accurate.  Field stars
were added randomly to approximately match the observed field star
density.  The artificial frames were then processed and measured in
exactly the same manner as the real data.

\subsubsection{The Artificial Cluster CMD}

In order to best understand the effects of crowding, we focus on the
G280 observations which have the largest contrast in surface brightness
between field and cluster.  G280 has one of the densest cores, is one of
the most extended clusters in our sample, and at the same time, has one
of the sparsest fields.  A280, the artificial analog of the G280
cluster, is shown in Figure \ref{fig:a280}.  This frame is composed of
450,000 cluster stars and 80,000 field stars.  The easiest way to
differentiate between the G280 (Fig. \ref{fig:g280}) and the A280
(Fig. \ref{fig:a280}) frames is to look for the presence (or absence) of
bad pixels, of which there are none on the artificial frame.

\begin{figure}
\epsscale{1}
\plotone{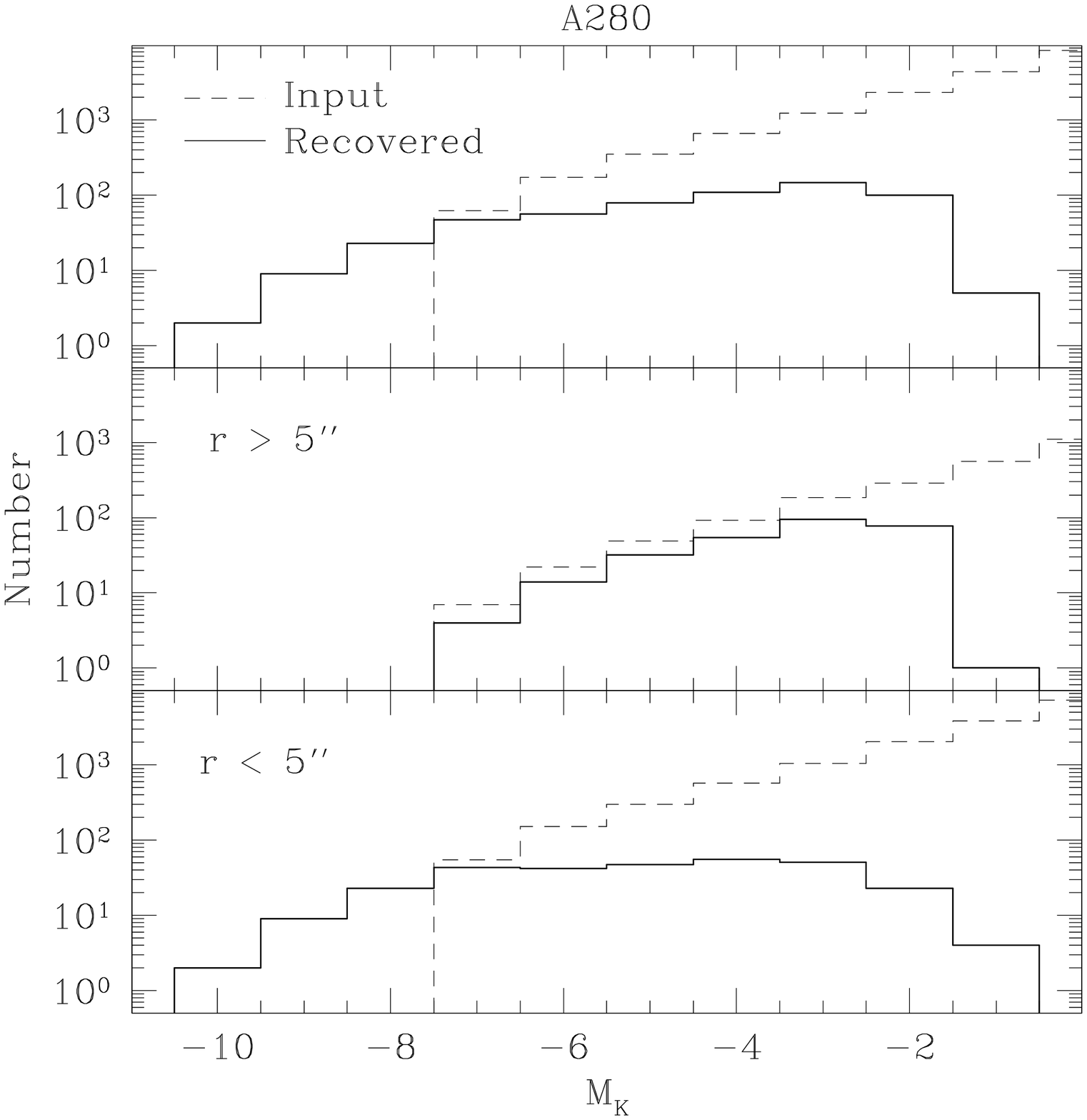}
\figcaption{
Artificial cluster A280 input (dashed) and recovered (solid) luminosity
functions.  Top: the entire frame.  Center: only stars located $>5''$
from the cluster center.  Bottom: stars found $<5''$ from the cluster
center.  The faint end decline is due to incompleteness, the extension
of the measured LF brighter than the input LF is due to blending.
\label{fig:a280lf}}
\end{figure}

Figure \ref{fig:a280lf} shows the input (dashed) and measured (solid)
luminosity functions for the A280 frame.  The top panel is for the whole
frame.  Even though the input LF cuts off at $M_K=-7.2$, the measured LF
extends to $M_K \sim -10$.  Assuming that most of the recovered stars
within $\sim 5''$ belong to the cluster, we break the LF up into the
cluster and field contributions.  The middle panel shows the LF for the
field around G280, only counting objects located greater than $5''$ from
the cluster center.  Here the measured LF shows very good correspondence
with the input LF, having identical bright end cutoffs.  The faint end
decline at $M_K \sim -2$ is due to ``classical'' incompleteness.  The
bottom panel includes only objects closer then $5''$ from the cluster
center; we assume that these are mostly cluster stars.  This is where
the major discrepancy between the input and recovered LF occurs.  All of
the stars measured brighter than the input LF cutoff, i.e. brighter than
any ``real'' star, are found in this small region of the frame.  We
emphasize that the same LF was used for the entire frame.

\begin{figure}
\epsscale{1}
\plotone{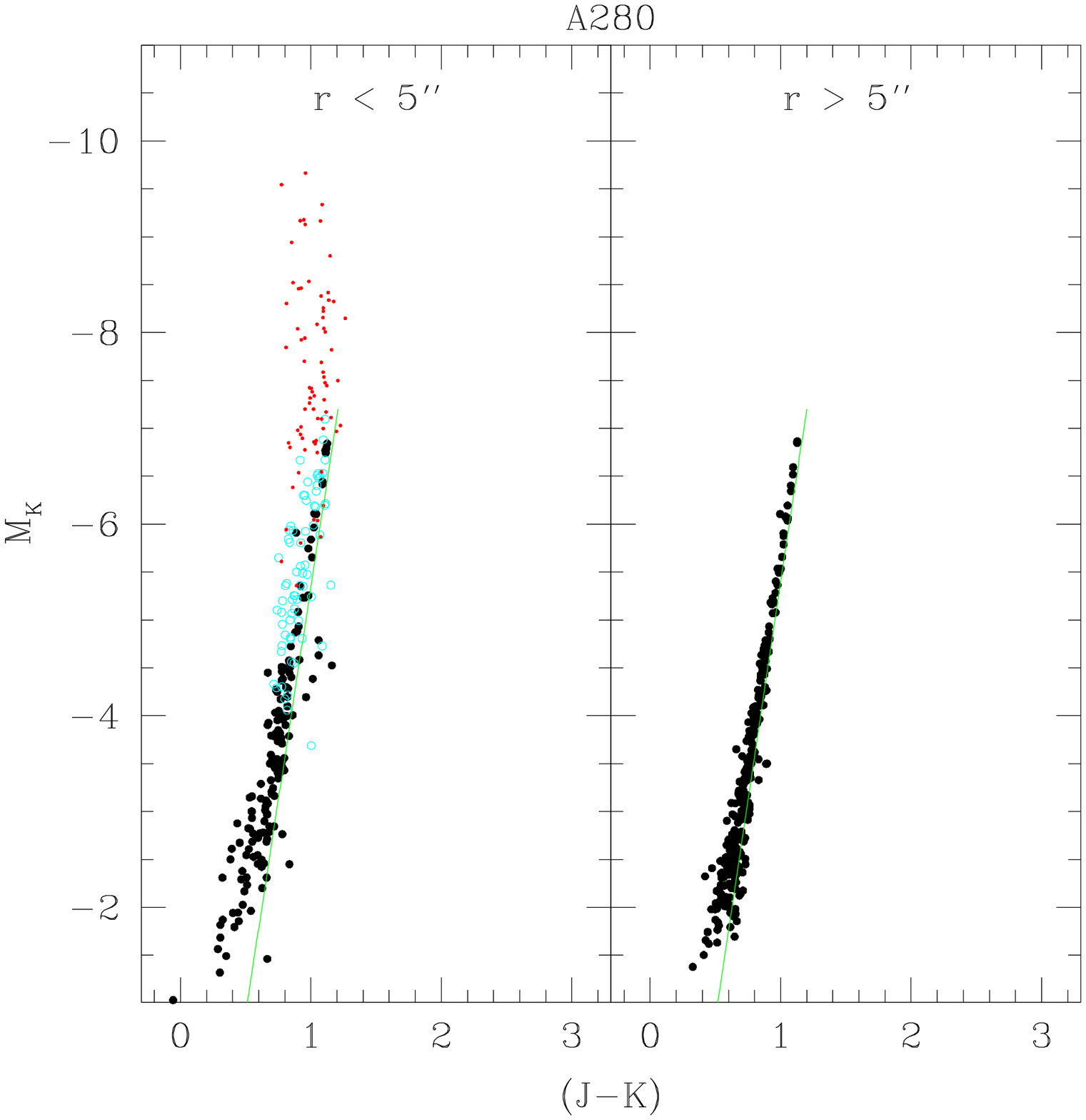}
\figcaption{
Artificial cluster A280 input (solid line) and recovered CMDs.  The left
panel shows objects measured within $5''$ of the cluster center, and the
right panel objects measured farther than $5''$.  Using criteria
developed in \S \ref{sec:results}, we plot objects within the
threshold-blending limit $(\mu_K<16, r<2.1'')$ as open circles, and
objects within the critical-blending limit $(\mu_K<14, r<1.3'')$ as
half-size dots.
\label{fig:a280cmd}}
\end{figure}

The A280 CMDs are shown in Figure \ref{fig:a280cmd} where we have again
divided the field spatially.  Objects closer than $5''$ from the cluster
center are plotted in the left panel, while objects farther than $5''$
are in the right panel.  Points and circles indicate measured objects;
the locus of the input stars is indicated by the line.  In the right
panel there is a very good correspondence between the input and
recovered magnitudes.  There is very little crowding, and as a result
very little scatter of the measured data around the input data.  The
scatter at the faint end gives an idea of the scale of the measurement
errors.  In the left panel, however, there is a large plume of stars
measured brighter and bluer than anything input into the cluster.  This
blue color is an immediate indicator of the blended origin of these
objects, where what is measured is not a single bright star, but the
superposition of many fainter, bluer stars.  Using criteria developed in
\S \ref{sec:results}, we plot objects inside the threshold-blending
radius, where the blending is just beginning to affect the faintest
stars, at surface brightness of $\mu_K=16$ magnitudes arcsecond$^{-2}$,
with open circles.  Objects measured inside the critical-blending
radius, where the surface brightness is $\mu_K \leq 14$ and blending has
severely distorted nearly all measurements, with half-size dots.  Note
that nearly all recovered stars lie systematically to the blue of the
input line for $r<5''$.  This suggests that blending is affecting all
stars to some degree, which is not surprising since $r \sim 5''$ is also
the dividing line inside which there were added more than one star per
pixel.

\begin{figure}
\epsscale{1}
\plotone{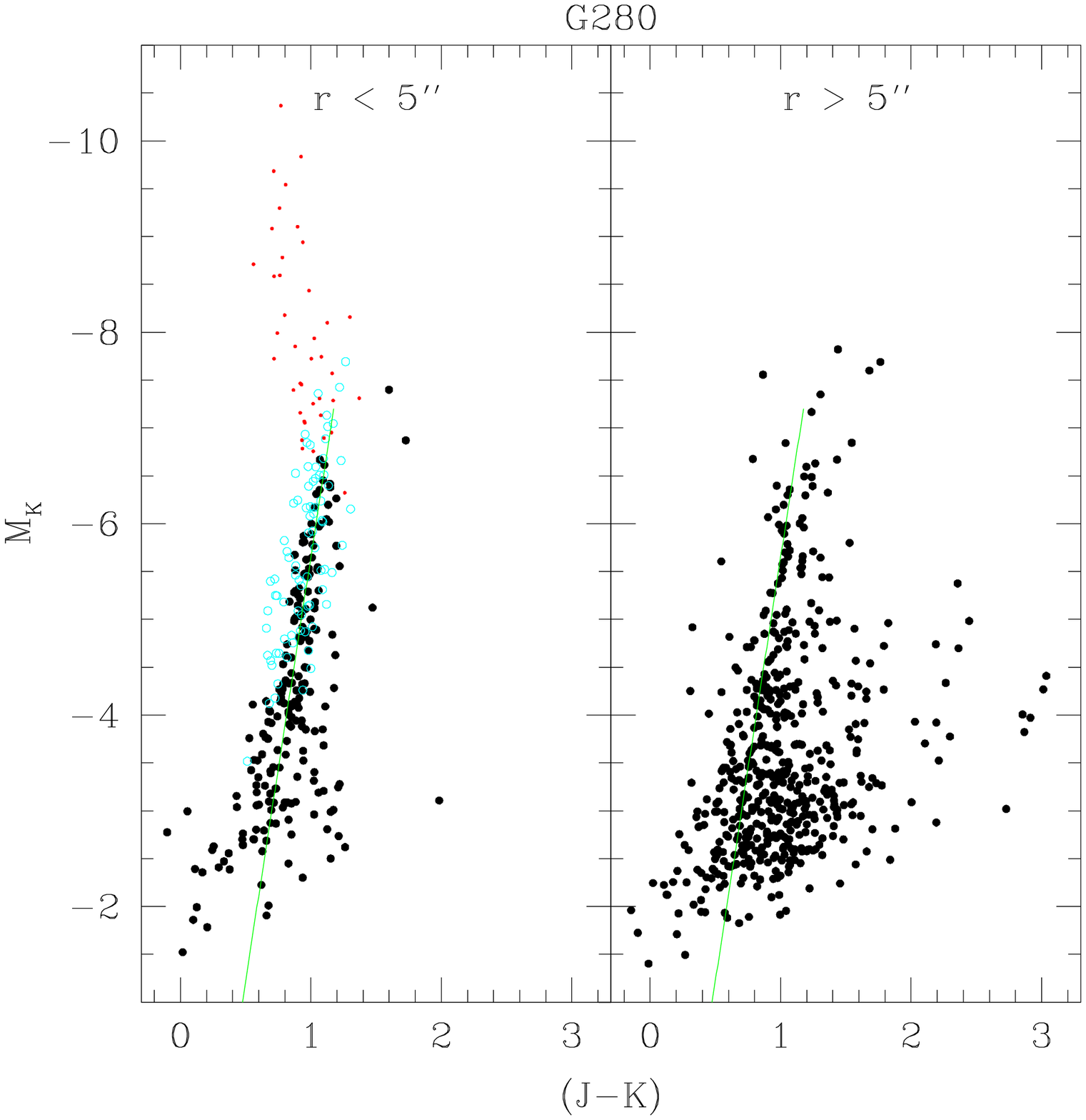}
\figcaption{
G280 -- Left: all objects measured within $5''$ of the center of G280.
Open circles indicate objects within the threshold-blending limit
$(\mu_K<16, r<2.2'')$, and half-size dots are for objects inside the
critical-blending limit $(\mu_K<14, r<1.0'')$.  Right: objects farther
than $5''$ from the cluster.  The solid line is the RGB measured in the
Galactic bulge (as well as the input into our artificial clusters).
\label{fig:g280cmd}}
\end{figure}

The G280 $M_K$-$(J-K)$ CMDs are shown in Figure \ref{fig:g280cmd} for
comparison.  The left panel shows all the data inside a radius of $5''$,
the distance chosen to define the cluster.  Objects inside the
threshold-blending limit $(\mu_K<16, r<2.2'')$ are plotted with open
circles, and objects inside the critical-blending limit $(\mu_K<14,
r<1.0''$) are plotted with half-size dots.  These radii are determined
in \S \ref{sec:results}, and listed in Table \ref{tab:blend_radii}.  The
right panel shows all objects outside the $5''$ cluster radius.  These
stars are expected to be non-cluster, or ``field'' stars.

The color spread in the G280 field CMD is real.  This field is $\sim
85\%$ disk and $\sim 15\%$ bulge, with an estimated spread in [Fe/H] of
$\sim 0.5$ (Paper II).  The artificial A280 frame, in contrast, was
created using a single metallicity, and thus shows a much smaller color
spread.  The field subtracted cluster CMD of G280 also looks remarkably
similar to the A280 cluster CMD (paper II).

Table \ref{tab:a280blends} summarizes the true blending of stars over a
range in measured luminosities in the A280 cluster.  We have randomly
chosen stars with approximately integer measured magnitudes, and assessed
their environment and number of neighbors which are contributing to their
measured flux.  The first three columns list the measured absolute
$K$-band magnitude, the distance of the star from the center of the
cluster, and the average surface brightness at that distance
respectively.  The fourth column lists the number of stars which were
input within a 2 pixel ($0.15''$; about the size of one resolution
element) radius of the measured star's position, and the last column the
number of stars brighter than $M_K=-2$ within 2 pixels.  Even though
there are {\em many} faint stars within 2 pixels of {\em most} of the
measured stars, the number of such stars which are relatively bright is
much less.

\begin{deluxetable}{ccccc}
\tablewidth{7.5cm}
\tablecaption{A280 Blends}
\tabletypesize{\footnotesize}
\tablehead{
\colhead{$M_K$}			&
\colhead{Radius}		&
\colhead{$\mu_K$}		&
\colhead{$N_{2pix}$}		&
\colhead{$N_{2pix}$}		\\
\colhead{}			&
\colhead{($''$)}		&
\colhead{}			&
\colhead{($M_K<5.8$)}		&
\colhead{($M_K<-2$)}		}
\startdata
-8.94	& 0.38	& 12.3	& 12777 & 88	\\
-8.01	& 0.73	& 13.4	&  4233	& 36	\\
-7.00	& 1.10	& 14.4	&  1661	& 11	\\
-5.99	& 1.53	& 15.5	&   644	&  6	\\
-5.00	& 1.71	& 15.9	&   451	&  5	\\
-4.01	& 2.61	& 17.7	&    87	&  1	\\
-3.01	& 9.48	& 20.2	&    10	&  1	\\
\enddata
\label{tab:a280blends}
\end{deluxetable}

\subsubsection{Effects of Varying the Input LF}

We have run simulations varying the shape of the LF input into the
artificial cluster.  The results are summarized in Table
\ref{tab:variable_lfs}.  The first two columns list the limits of the
power-law input LF, and the third column is the slope $\alpha=0.278$
($N(m)\propto 10^{\alpha m}$).  The fourth column gives the number of
stars input into the cluster; this number was scaled in order to keep
approximately the same integrated cluster magnitude.  The last two
columns list the brightest star measured, and the average $(J-K)$ color
of the five brightest stars in each cluster.  Note that the ``normal''
A280 LF is taken from Baade's Window with a slope of 0.278 extending
from $5.8<M_K<-7.2$, and has 450,000 stars.

The top half of Table \ref{tab:variable_lfs} describes the simulations
where we varied the bright end extent of the input LF by $\pm 2$
magnitudes.  All the simulations achieve approximately the same core
surface brightness, and, as a result, generate blended objects which all
reach nearly the same brightness.  The most noticeable difference
between these simulations is the recovered colors.  The simulation with
the faintest input LF cutoff (and hence the most stars), returns stars
with bluer colors, by as much as $(J-K)\sim 0.4$ at the bright end.

The bottom half of Table \ref{tab:variable_lfs} is for the simulations
where we varied the input LF slope by $\pm 0.04$.  As before, all
simulations achieve approximately the same core surface brightness, and
blends up to $M_K \sim -9.7$.  The $\alpha = 0.318$ trial has a
noticeably wider RGB due to the many blends with fainter stars.
However, the brightest blends do not show the same gradual progression
towards bluer colors with increased number of input stars as was
observed when varying the input LF cutoff.

Can these simulations recover the true LF from a blended frame?  In the
fields surrounding the clusters, the simulations confirm that what we
measure is indeed the true LF.  However, in the cores of the clusters
where the stars are severely blended, there is a degeneracy between the
combination of LF extent, slope, and number of stars (surface
brightness).  In these dense regions it is nearly impossible to
distinguish between a cluster with 250,000 stars going as bright at $M_K
\sim -9$, and another cluster with stars as bright as $M_K \sim -5$ but
three times as many stars, at least using the current observations and
techniques.

\begin{deluxetable}{cccccc}
\tablewidth{8.5cm}
\tablecaption{A280 Variable LFs}
\tabletypesize{\footnotesize}
\tablehead{
\colhead{$M_{K1}$}		&
\colhead{$M_{K2}$}		&
\colhead{$\alpha$}		&
\colhead{$N_{input}$}		&
\colhead{$M_K$}			&
\colhead{$(J-K)$}		\\
\multispan2{(input limits)}	&
\colhead{}			&
\colhead{}			&
\colhead{(brightest)}		&
\colhead{(Top5)}		}
\startdata
5.8  & -9.2  & 0.278  & 254000 & -9.85  & 1.11  \\
5.8  & -8.2  & 0.278  & 338000 & -9.92  & 1.03  \\
5.8  & -7.2  & 0.278  & 450000 & -9.66  & 0.94  \\
5.8  & -6.2  & 0.278  & 601000 & -9.93  & 0.87  \\
5.8  & -5.2  & 0.278  & 805000 & -9.41  & 0.72  \\ 
\tableline
5.8  & -7.2  & 0.238  & 207000 & -9.74  & 1.12  \\
5.8  & -7.2  & 0.258  & 306000 & -9.77  & 0.93  \\
5.8  & -7.2  & 0.278  & 450000 & -9.66  & 0.94  \\
5.8  & -7.2  & 0.298  & 653000 & -9.61  & 0.93  \\
5.8  & -7.2  & 0.318  & 933000 & -9.69  & 0.93
\enddata
\label{tab:variable_lfs}
\end{deluxetable}

%
%
\subsection{Comparison of Artificial Star Tests} \label{sec:comparison}

We now compare the luminosity functions from the traditional
completeness tests, where a handful of artificial stars were added to an
observed frame, to those from the artificial cluster tests.

\begin{figure}
\epsscale{1}
\plotone{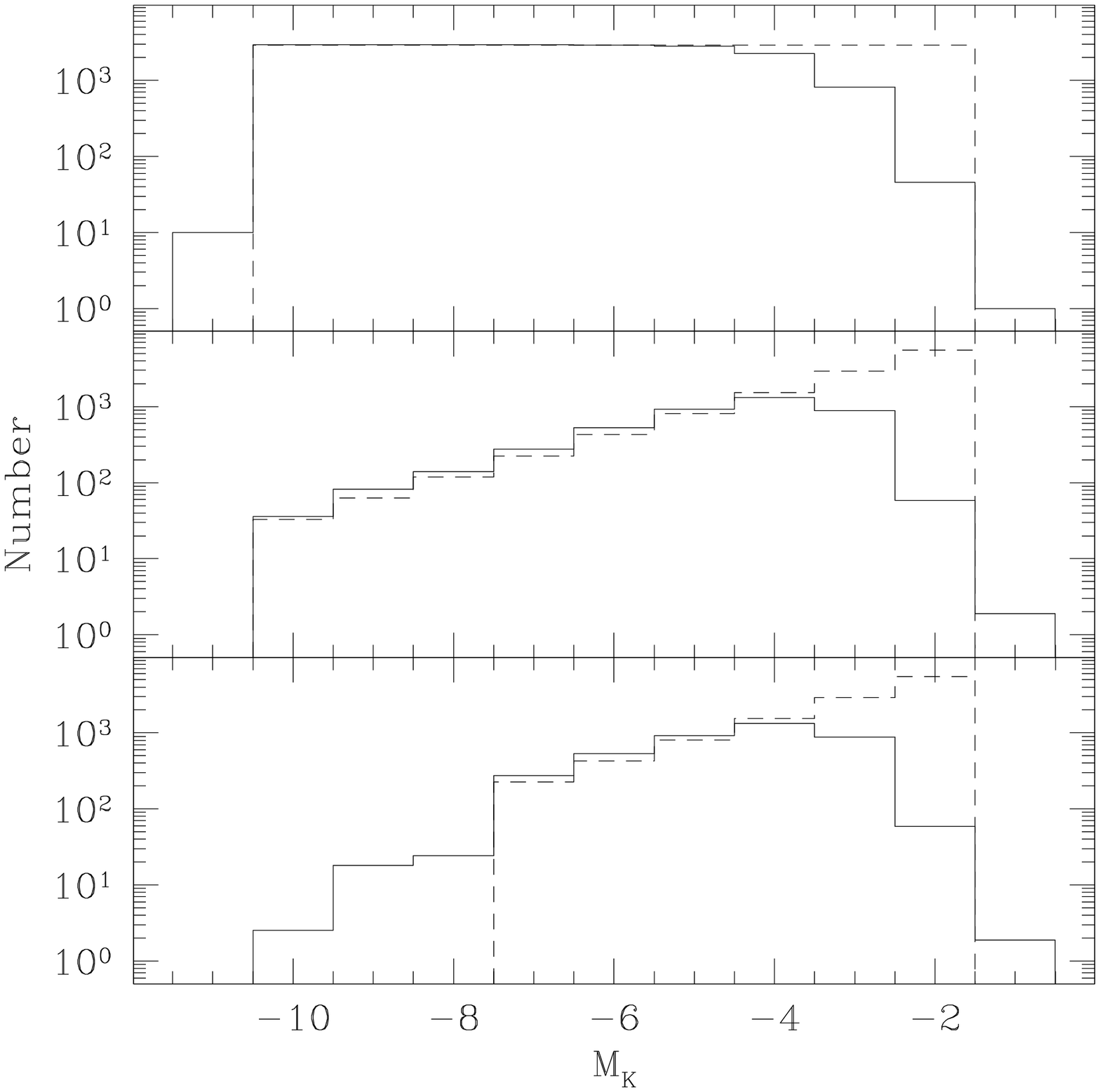}
\figcaption{Transformation of the traditional completeness test results
to mimic the power-law input LF used in the artificial cluster tests.
Top: initial boxcar input LF (dashed) and the recovered LF (solid).
Center: input LF multiplied by a power-law and the recovered LF.
Bottom: input LF truncated at $M_K=-7.3$ (plotted with 1 mag wide bins)
and the resulting recovered LF.  Note that all stars measured with
$M_K<-7.3$ are not real.
\label{fig:artstarlf}}
\end{figure}

We start with the traditional completeness test LF shown in the top
panel of Figure \ref{fig:artstarlf}.  The dashed line illustrates the
input LF composed of 2916 input stars per magnitude, achieved 81 stars
at a time over 36 trials.  The solid line is the recovered LF.  To mimic
a power-law input LF, we multiply the input LF by the same power-law
slope as was used in the artificial cluster tests ($\alpha = 0.278$).
The number of stars recovered (at any magnitude) is then multiplied by
the same constant that was applied to the responsible input magnitude.
The recovered stars from each input magnitude are then summed to get the
total recovered LF.  The resulting input (dashed) and recovered (solid)
LFs are shown in the middle panel of Figure \ref{fig:artstarlf}.

However, from our initial completeness tests,
we input stars several magnitudes brighter than the true extent of the LF.
(This is because with just the traditional completeness tests,
the true brightnesses of the measured stars are unknown,
because one is just placing artificial stars on the preexisting clumps of stars.)
Thus we now truncate the input LF at $M_K=-7.3$ to (nearly) match
the artificial cluster input LF.
The results are shown in the bottom panel of Fig. \ref{fig:artstarlf}.

The traditional completeness test results (bottom
Fig. \ref{fig:artstarlf}) are approximately the same as the artificial
cluster results (top Fig. \ref{fig:a280lf}).  The input LFs (dashed
lines) start at $M_K \sim -7$ and follow a power-law distribution upward
toward fainter magnitudes.  The observed LF (solid line) extends three
magnitudes brighter than the input LF, and drop off at $M_K \gtrsim -2$
due to incompleteness.

At the faint end of the photometry, both tests are equally useful to
characterize the photometric completeness.  The artificial clusters
simulations are, however, a more controlled experiment that offer
advantages to the interpretation of the nature of the brightest measured
stars.  The traditional completeness tests show that some stars are
brightened, but when the crowding gets severe, the brightening seems to
come from falling on or near a preexisting bright clump, and measuring
the brightness of the clump rather than the individual star.  These
tests, however, do not tell you whether the bright clumps are real.  The
artificial cluster simulations, on the other hand, tell you the number
and brightness of the stars which make up the clumps, and in our case,
that these bright objects which are observed can be created with a
normal LF whose brightest star is several magnitudes fainter than the
brightest object observed.

%
%
\subsection{Artificial Fields} \label{sec:artificial_fields}

We have created uniformly populated mini-fields of $100 \times 100$
pixels to investigate any possible effects of the strong surface
brightness gradient present in the observed globular clusters.  This is
especially important since we use surface brightness as the criterion
for quantifying blending, and determining the believability of our
photometry.  Each mini-field is constructed in the same manner as the
artificial clusters, using a power-law input luminosity function,
extending across $-7.2 < M_K < 5.8$.  Between $10^4$ and $\sim 5 \times
10^6$ stars are randomly added to each field, where the field with the
most stars nearly reaches the surface brightness level observed in the
cores of the M31 clusters.

\begin{figure}
\epsscale{1.25}
\plotone{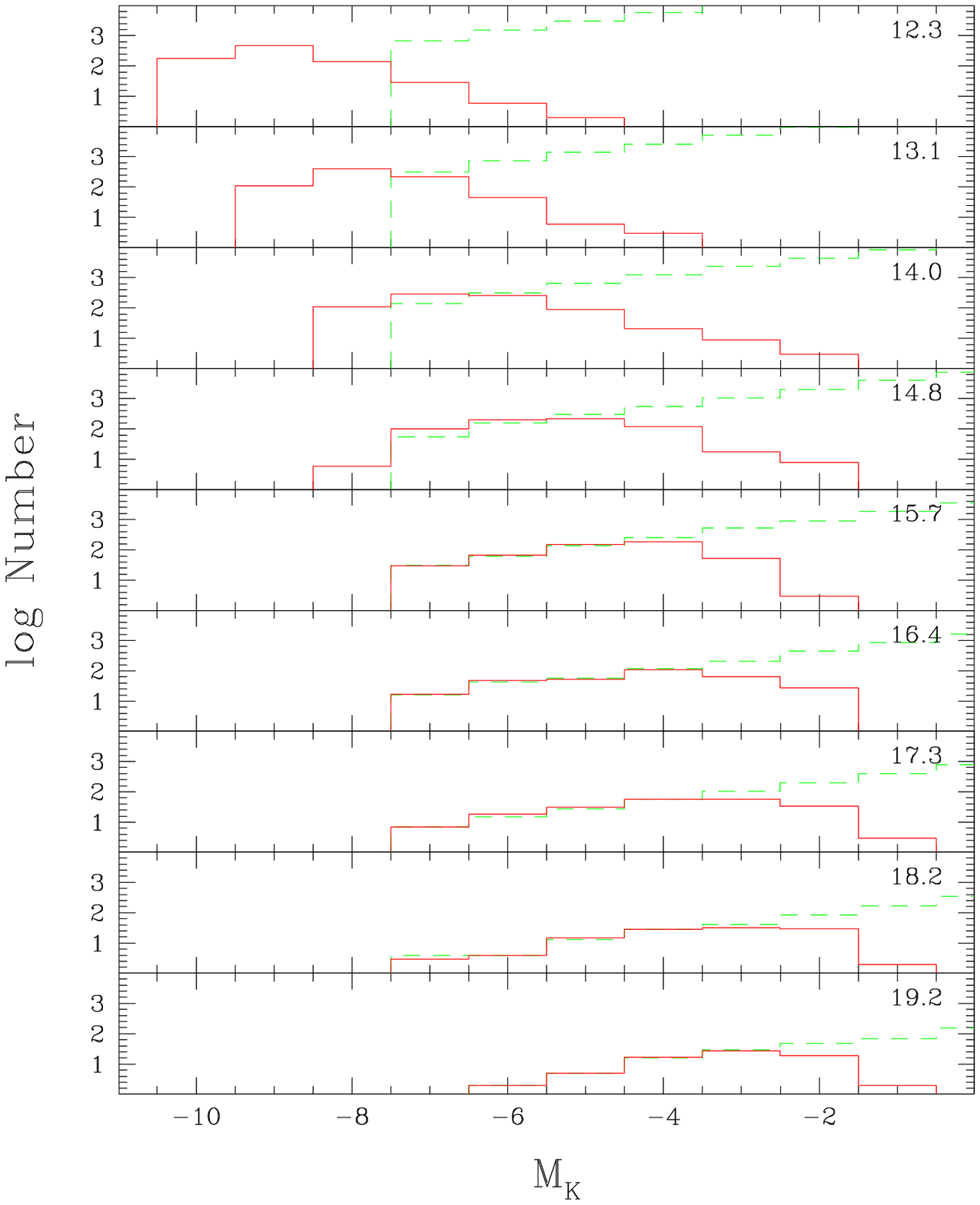}
\figcaption{
Input (dashed) and recovered (solid) luminosity functions of artificial
fields of varying surface brightnesses.  All fields have the same input
LF slope, and the average surface brightness of each field is listed in
the upper right corner of each panel.  The brightest star input into any
field is $M_K = -7.2$ (binned into 1 magnitude wide bins), however due
to small number statistics, the less-populated fields may not have any
stars this bright.  Note how the recovered LF not only shifts toward
brighter magnitudes with increasing surface brightness, but also changes
slope due to the effects of blending.
\label{fig:artfieldlfs}}
\end{figure}

The input (dashed lines) and recovered (solid lines) luminosity
functions of these fields are shown in Figure \ref{fig:artfieldlfs}.
The surface brightness of each field is listed in the upper right corner
of each plot.  There is good agreement between the bright end extent of
the input and measured LFs up to a surface brightness of $\mu_K=15.7$.
However as more and more stars are added to the field, blending becomes
significant, the LF shifts toward brighter and brighter magnitudes, and
then more stars are ``recovered'' with $M_K$ brighter than the brightest
input star.

Notice the change of the slope of the measured LF.  When blending is
insignificant, the measured LF is very close the the input LF,
i.e. increasing toward fainter magnitudes with the correct slope.
However when blending becomes severe, more stars clump together, and
fewer faint stars are recovered.  The relative number of bright stars
compared to faint stars changes dramatically, and the slope of recovered
LF now has an opposite sign compared to the input LF.

Our tests show that the surface brightness gradient appears
inconsequential, insofar as the effects of blending on photometry are
concerned.  An advantage of the mini-fields over the artificial clusters
is that they give a larger sample of stars at each surface brightness
interval.  This makes it easier to quantify at what surface brightness
levels the photometry is either accurate, tainted, or corrupt.

%
%
\section{Results} \label{sec:results}

We find that many of the bright objects observed in crowded regions
arise from the clumping of many fainter stars.  Although the traditional
completeness tests do show that faint stars are artificially brightened,
they do not give the whole picture.  Our artificial frames, which
closely mimic the real HST-NICMOS data, have shown that random clumps of
stars can create objects several magnitudes brighter than the brightest
real stars.

\begin{figure}
\epsscale{1}
\plotone{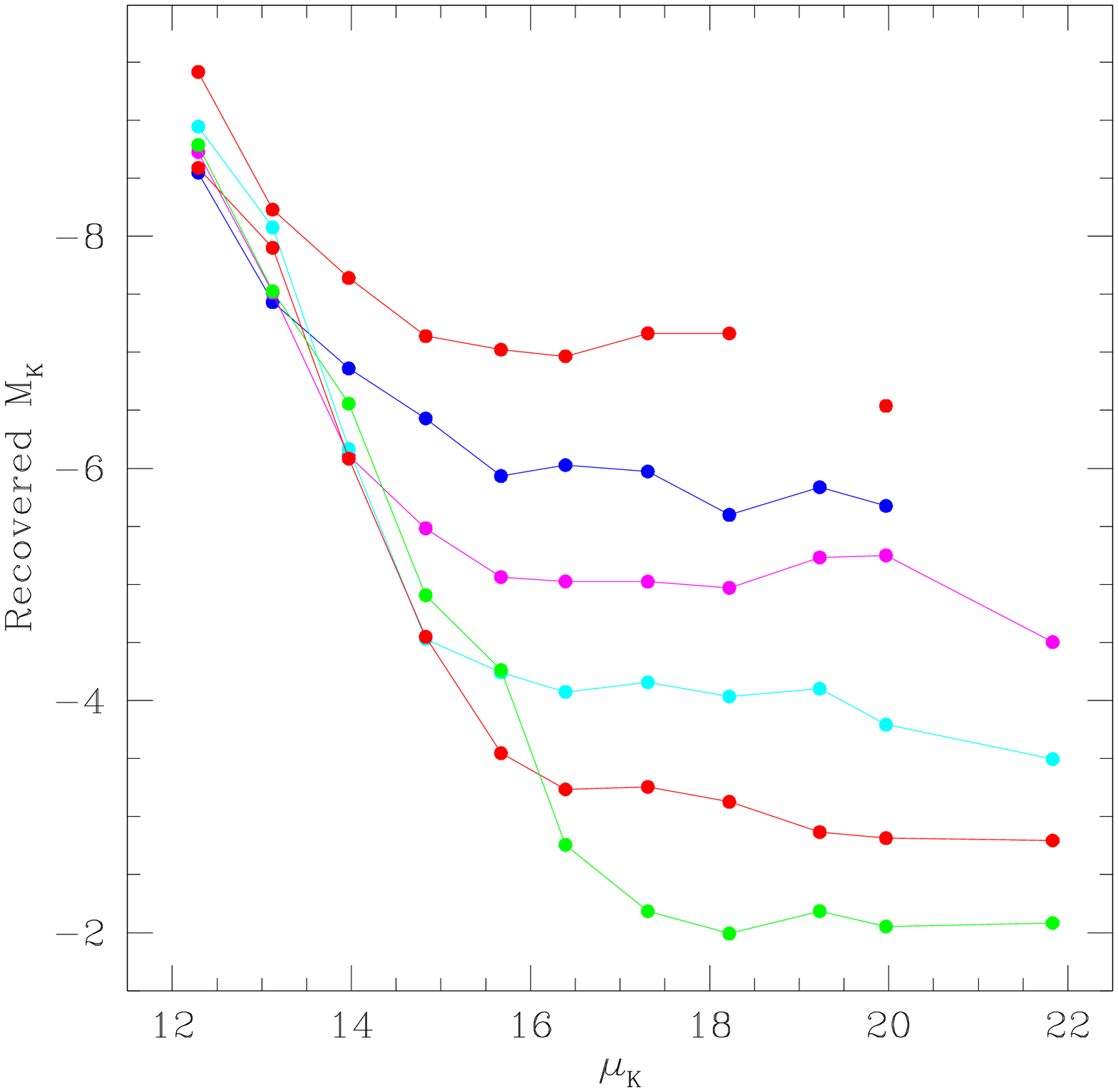}
\figcaption{
The recovered magnitude of stars measured in mini-fields over a range of
surface brightnesses.  Each line indicates the median recovered
magnitude over a range of $\pm 0.5$ magnitudes in input brightnesses.
The input magnitude can be ascertained from the measurements made at low
surface brightnesses which are accurate, but affected by small number
statistics.  Thus one can monitor the effects of crowding on the
measurements of stars of $\sim$ constant magnitude.  Note the similarity
of this plot to Fig. \ref{fig:artstarfits}.
\label{fig:artfieldsb}}
\end{figure}

Figure \ref{fig:artfieldsb}, a plot of the recovered absolute $K$-band
magnitude as a function of the field surface brightness, shows the
results of this blending.  Here each curve represents the median of
stars input with a range of $\pm 0.5$ magnitudes.  For example the
bottom-most curve gives the median recovered $M_K$ of all stars input
with magnitudes between $-1.5$ and $-2.5$ in each $100 \times 100$ pixel
mini-field (7\farcs6 $\times$ 7\farcs6).  The number of stars in each 1
magnitude range is a function of the total number of stars input
(i.e. the surface brightness), and the range being considered, since we
use the same power-law input luminosity function.  Thus the statistical
noise is quite high at the faint surface brightness end, especially for
the brighter input curves.  Note that there were no stars at $M_K=-6$ or
$M_K=-7$ at $\mu_K \sim 22$, or for $M_K = -7$ at $\mu_K \sim 19$.

\begin{figure}
\figurenum{9b}
\epsscale{1}
\plotone{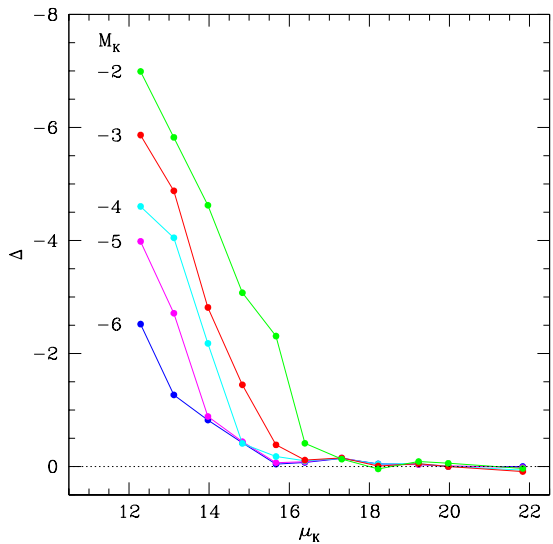}
\figcaption{
The median difference between the input and recovered magnitudes of
stars measured in mini-fields of different average surface brightness.
Each line indicates a single range of input magnitudes, which are
labeled at the leftmost termination of each.
\label{fig:artfieldsb_deltamag}}
\end{figure}

Figure \ref{fig:artfieldsb_deltamag} shows a plot of the average
deviation between the input and recovered magnitudes for stars of
different input brightnesses, as a function of the average surface
brightness of the artificial field.  This plot illustrates how
increasing surface brightness has an increasingly strong effect on
stellar photometry.  At surface brightnesses fainter than $\mu_K \sim
17$ magnitudes arcsecond$^{-2}$, all stars are measured fairly
accurately.  However at $\mu_K \sim 16$ the fainter stars start to
become artificially brightened by crowding.  With increasing surface
brightness, the effects of crowding become more and more significant for
brighter stars.  At $\mu_K \lesssim 13$ the crowding is so severe that
nearly all stars are badly blended or misidentified with random clumps
of blended stars, and even the brightest input stars are mismeasured by
several magnitudes.

Comparing the fits to the recovered artificial star magnitudes from the
traditional completeness tests (Figure \ref{fig:completeness}), and the
recovered magnitudes from the artificial mini-fields (Figure
\ref{fig:artfieldsb_deltamag}), the same effect is apparent.  At some
critical surface brightness, or radius in the case of our globular
clusters, blending effects begin to dominate.  As the surface brightness
increases, blending becomes more and more important, until even the
brightest stars become lost in the uneven background (for more on the
effects of crowding on positional errors see \citet{Hog2000}).

What traditional completeness tests (adding stars to an existing frame),
do not show is the true upper limit to the LF.  For example, look at the
plot of recovered magnitudes from the artificial mini-fields (Figure
\ref{fig:artfieldsb}).  Suppose an object is measured at $M_K \sim -6$
in a region where the surface brightness is $\mu_K \sim 14$.  This
object could be a star of $M_K \sim -6$, or it could just as easily be a
blend of many stars several magnitudes fainter.

Using Figures \ref{fig:artfieldsb} and \ref{fig:artfieldsb_deltamag} as
guides, we chose $\mu_K = 16$ as the threshold point where our
photometry starts to become noticeably affected by blending, and $\mu_K
= 14$ as the critical blending radius where our photometry is dominated
by blends.  Beyond this level, no measurements are reliable.

\begin{figure}
\epsscale{1}
\plotone{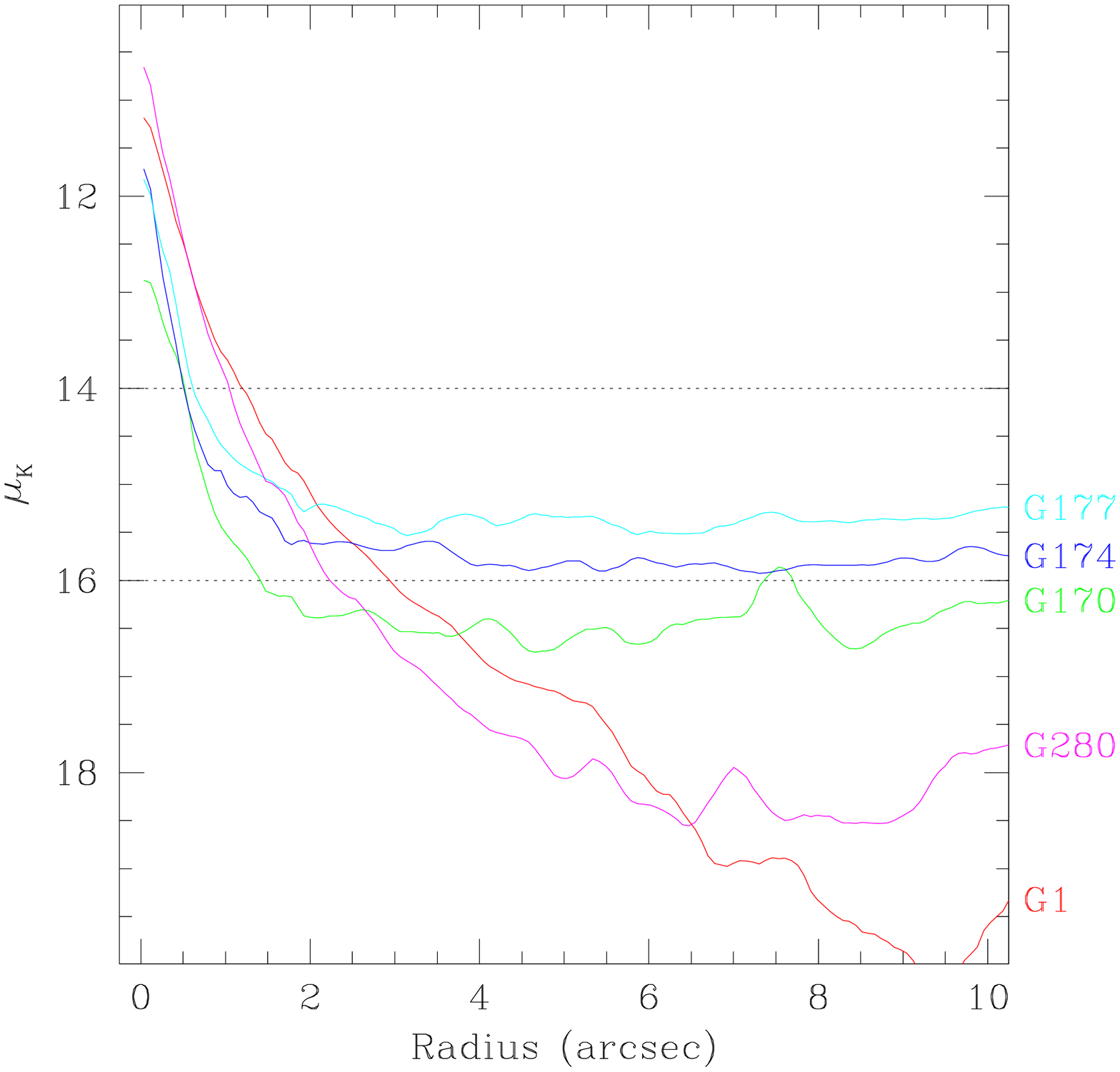}
\figcaption{
Azimuthally averaged $K$-band surface brightness as a function of radius
from the cluster center.  The dashed lines show the threshold--
$(\mu_K=16)$ and critical--$(\mu_K=14)$ blending surface brightnesses .
\label{fig:surface_brightness}}
\end{figure}

Figure \ref{fig:surface_brightness} shows azimuthally averaged surface
brightness profiles of each cluster (surface brightness values have been
normalized from \citet{Ken1987} photometry as described in \S
\ref{sec:data_reduction}).  Dotted lines indicate the threshold- and
critical-blending surface brightness levels.  We use this plot to
determine the threshold-blending radius ($R_{16}$), and the
critical-blending radius ($R_{14}$) for each cluster.  These radii are
listed in Table \ref{tab:blend_radii}.  Any objects measured inside the
threshold-blending radius, especially faint objects, are potentially
affected by blending, and should be considered suspect.  (Note that this
radius ($R_{16}$) was chosen so that stars input at $M_K \sim -3$ could
be recovered accurately most of the time.  However, for stars fainter
than this, there is no way to tell whether stars measured at $M_K > -3$
are blends or not.)  Objects measured inside the critical-blending
radius are undoubtedly blends, and although we plot them for
completeness, they should be disregarded.

\begin{deluxetable}{cccc}
\tablewidth{5cm}
\tablecaption{Threshold \& Critical Blend Radii}
\tabletypesize{\footnotesize}
\tablehead{
\colhead{Name}		&
\colhead{Core $\mu_K$}	&
\colhead{$R_{14}$}	&
\colhead{$R_{16}$}	}
\startdata
G001    & 11.2  & 1.2   & 2.9           \\
G170    & 12.9  & 0.5   & 1.4           \\
G174    & 11.7  & 0.5   & \nodata       \\
G177    & 11.8  & 0.6   & \nodata       \\
G280    & 10.7  & 1.0   & 2.2 
\enddata
\label{tab:blend_radii}
\end{deluxetable}

In summary, traditional completeness tests are good for analyzing one's
detection efficiency and photometric errors.  However, in the case of
blended images, these tests cannot reveal what the blends are composed
of.  Only by creating completely artificial frames, and matching the
observed and modeled surface brightnesses, can one get an idea of which
stellar populations are consistent with the observations.  Even then
there are degeneracies between the extent of the LF, the slope of the
LF, and the true number of stars.

%
%
\subsection{Effects of Blending on the RGB Slope and Width} \label{sec:effects_of_blending}

Since in Paper II we will use the slope of the RGB to estimate
metallicities, and the width of the RGB to place limits on any spread in
metallicity, we have investigated the effects of crowding on these
quantities.  This specific case should serve as a good example for
other, similar analyses.  This analysis uses the mini-fields discussed
in the previous section, which include between $10^4$ and $\sim 5 \times
10^6$ stars, achieving surface brightnesses in the range $19.2 \lesssim
\mu_K \lesssim 12.3$.  The input $M_K$-$(J-K)$ RGB slope is $-0.113$,
which implies [Fe/H] $\sim -0.29$ according to the RGB slope --
metallicity relation of \citet{KF1995}.

We perform an iterative linear least squares fit to the measured RGB,
rejecting measurements farther than $\pm 3 \sigma$ from the fit.  Even
though the metallicity relation was derived for the RGB in the range $-1
\gtrsim M_K \gtrsim -6$, here we fit to all the data since blending
shifts the RGB several magnitudes in luminosity.

\begin{figure}
\epsscale{1}
\plotone{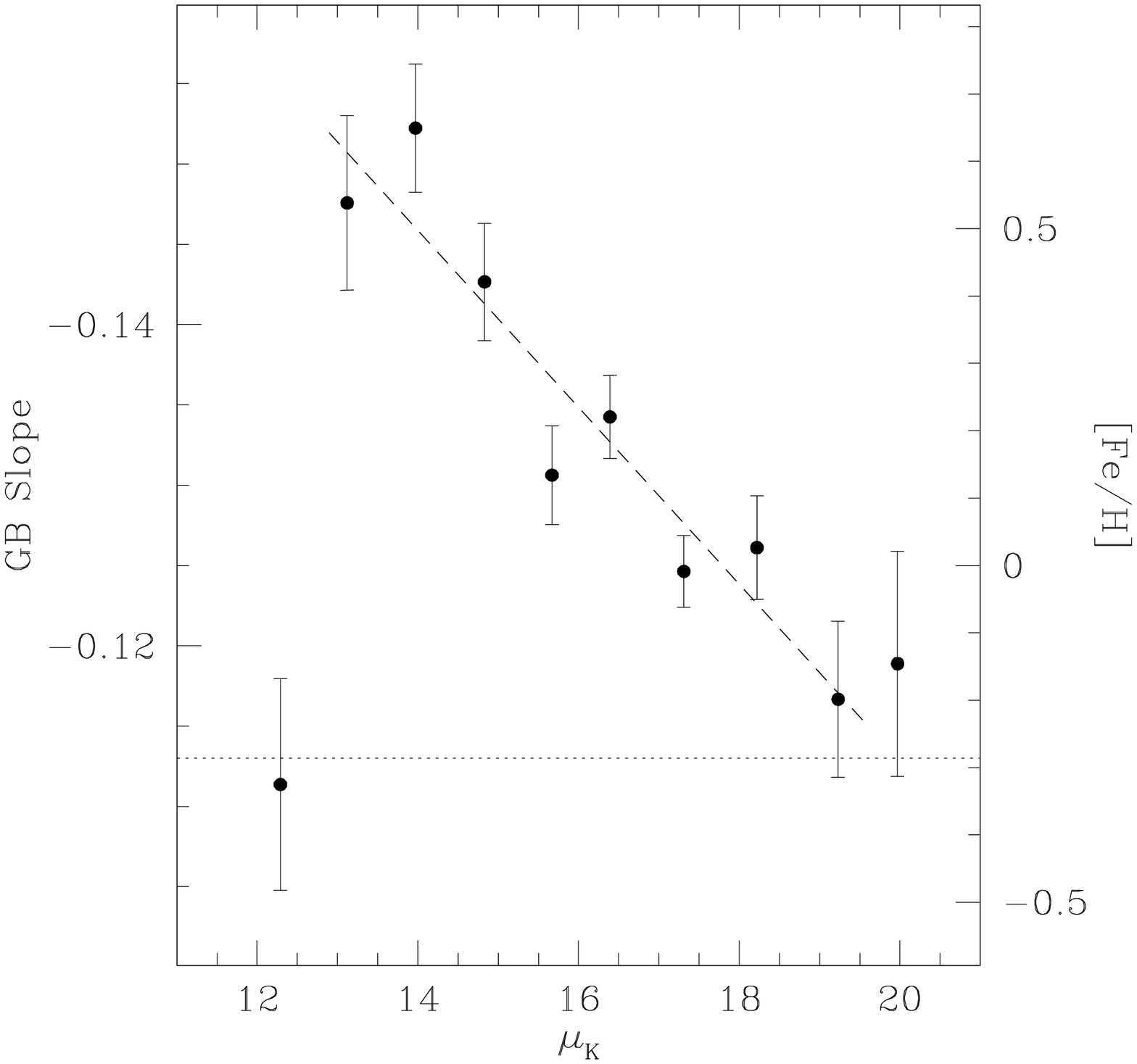}
\figcaption{Derived giant branch slope and the implied metallicity 
as a function of field surface brightness.  The dotted line shows the
input RGB slope of -0.113 ([Fe/H]=-0.29).  The dashed line is a linear
fit to data with $\mu_K > 13$.
\label{fig:gbslope}}
\end{figure}

Figure \ref{fig:gbslope} shows the RGB slope measured in each simulated
mini-field as a function of the average field surface brightness.  The
dotted horizontal line at a slope of $-0.113$ shows the input RGB slope,
i.e. the slope which should be recovered in every field if crowding were
unimportant.  Indeed at the low surface brightness end, the recovered
RGB slope approaches the input RGB slope.  As the surface brightness of
the field increases $(\mu_K \gtrsim 13)$, more faint blue stars clump
together pushing the lower end of the RGB bluer and to higher
luminosities, thus tilting the RGB which implies higher metallicities.
At even higher surface brightnesses $(\mu_K \lesssim 13)$, the blending
now also significantly affects even the brightest stars.  All the faint
stars are lost in the background, and the only objects measured are just
the brightest stars clumped together with many fainter bluer stars.

The dashed line is a linear fit to the data with $\mu_K > 13$, and is
the relation we use to estimate corrections to our measured RGB slopes.
We emphasize that this correction is merely an estimate of the effects
of crowding.  We do not plot the correction as a function of delta slope
or delta [Fe/H], in order to discourage blind use, as the effects of
blending will be different depending on the specific input luminosity
function and colors.

On the right side of Figure \ref{fig:gbslope} is the metallicity scale
implied from the relation of \citet{KF1995}.  This shows the strong
dependence of the metallicity on the slope, and thus the severe effect
that even minimal blending can have on the metallicity derived from the
RGB slope.

\begin{figure}
\epsscale{1}
\plotone{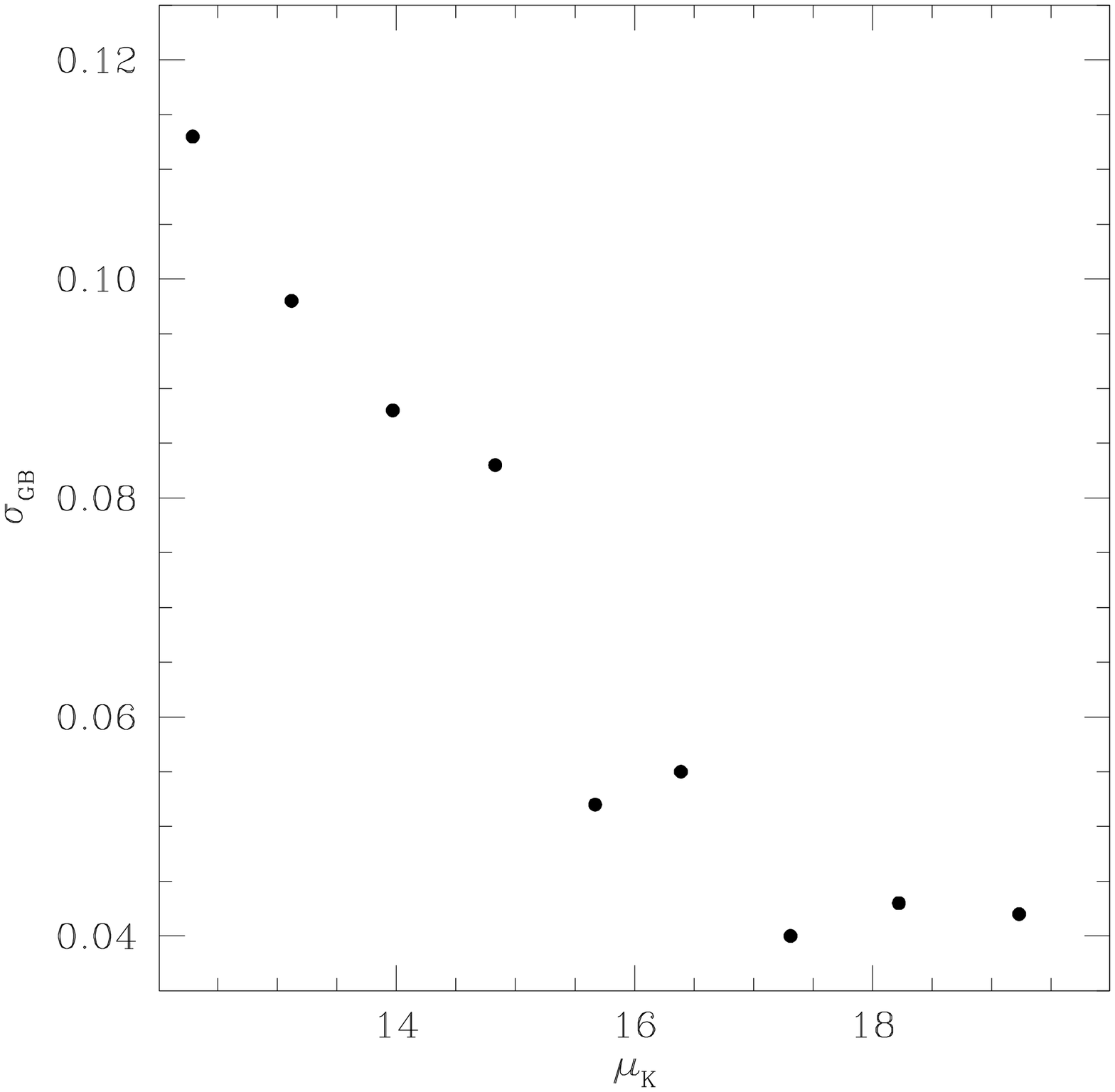}
\figcaption{Giant branch width as a function of the field surface brightness
\label{fig:gbwidth}}
\end{figure}

We now analyze the effects of blending on the width of the measured RGB.
All of our simulations start out with an input RGB of zero width.  To
measure the width of the RGB, the linear fit to the RGB is subtracted
off to leave a vertical RGB centered at zero $(J-K)$ color, and the
distribution is binned into 0.05 magnitude color bins.  We then fit a
Gaussian to the binned distribution.  Figure \ref{fig:gbwidth} shows the
width of the Gaussian fit to the measured giant branch of each field as
a function of the field's average surface brightness.  This plot shows
that the measured width of the RGB is not significantly affected by
crowding until the surface brightnesses exceeds $\mu_K \sim 17$.  At
that point the width grows approximately linearly with increasing
surface brightness.

%
%
\subsection{Simulations vs Theory} \label{sec:simulations_vs_theory}

A great deal on what to expect from crowding can be derived from first
principles \citep{Ren1998}. In this section the main points in this
theoretical approach are summarized and then compared to the results of
our simulations. For a given stellar population of age $t$ and total
bolometric luminosity $L_{\rm T}$ the number $N_{\rm j}$ of stars in the
generic $j$ (post main sequence) evolutionary stage of duration $t_{\rm
j}$ is given by
\begin{equation} 
N_{\rm j}=B(t)L_{\rm T}t_{\rm j}.  
\end{equation} 
where $B(t)$ is the specific evolutionary flux, i.e. the number of stars
leaving or entering any evolutionary stage per year and per unit (solar)
luminosity of the population. In practice, $B(t)$ is only weakly
dependent on age, varying from $\sim 0.5\times 10^{-11}$ to $\sim
2\times 10^{-11}$ \citep[stars \lsun$^{-1}$yr$^{-1}$, see Fig. 1
in][]{Ren1998}.  Equation (1) can be applied to one individual
resolution element, in which case the sampled luminosity can be written
as
\begin{equation} 
L_{\rm T,re} = s^2 \Sigma_{\rm T} = 
s^2 \Sigma_{{\rm T},\circ} \left( \frac{D}{200\,{\rm kpc}}\right)^2,
\end{equation}
where $s$ is the linear size of the resolution element (in arcsec),
$\Sigma_{\rm T}$ is the surface brightness of the stellar system 
in units of \lsun$/{\rm arcsec}^2$, $\Sigma_{{\rm T},\circ}$ is the surface 
brightness in units of \lsun$/{\rm pc}^2$, and $D$ is the distance.

For $N_{\rm j,re} = B(t) L_{\rm T,re} t_{\rm j} <1$, $N_{\rm j,re}$ is close
to the probability that a generic resolution element contains one star
in the $j$ phase. Therefore, $N_{\rm j,re}^2$ is the probability that a
resolution element contains a blend of two such stars that a photometric
package will mistake for a single star about twice as bright. More
generally, the probability that a resolution element will contain a
blend of a star in phase $j$ and a star in phase $k$ is given by:
\begin{equation} 
N_{\rm j,re}N_{\rm k,re} = B(t)^2 t_{\rm j} t_{\rm k} s^4
\Sigma_{{\rm T},\circ}^2 \left( \frac{D}{200\,{\rm kpc}} \right)^4.
\end{equation}
The bottom line is that, for given physical crowding as measured by the
surface brightness (in \lsun /pc$^2$ units), the number of two-star
blends is proportional to the fourth power of the linear size of the
resolution element, to the fourth power of the distance from the
observer to the stellar population, and to the square power of the
surface brightness \citep{Ren1998}.  For example, suppose we have
observed a globular cluster at a distance of 7 kpc with a resolution of
$1''$, and would like to obtain the same frequency of two-star blends
for an identical cluster in M31, i.e. at a distance of 700 kpc.
Equation (3) tells that we would need a resolution 100 times better,
i.e.  $\sim 0.01''$.

For example, Equations (2) and (3) can be used to estimate the average
number of RGB stars ($N_{\rm RGB}$), and RGB stars within one magnitude
of the RGB tip ($N_{\rm RGBT}$), per resolution element.  These can then
be used to determine the number of resolution elements in a frame that
contain a blend of two RGB stars ($N_{\rm 2RGB}$), two RGBT stars
($N_{\rm 2RGBT}$), three RGBT stars ($N_{\rm 3RGBT}$), etc.  Note that
the number of triplets scales as the cube of $N_{\rm RGBT}$, that of
quadruplets as its fourth power, and so on \citep{Ren1998}.

The results of our calculations are listed in Table
\ref{tab:theoretical_populations}.  Here we assume a 15Gyr population of
solar metallicity ($L_T = 0.36 L_K$), a distance modulus of 24.4, no
$K$-band extinction, a 376 arcsec$^2$ field of view, and we use the
$H$-band PSF FWHM of $0.15''$ as the size of our resolution element.

The first column is the $K$-band surface brightness ($\mu_K$), and the
second is the total bolometric luminosity sampled by a single resolution
element ($L_T$).  The third and fourth columns give the average number
of stars on the red giant branch ($N_{RGB}$) and stars within one
magnitude of the red giant branch tip ($N_{RGBT}$) sampled by a single
resolution element.  The fifth, sixth, seventh and eighth columns give
the number of 2 RGBT star blends ($(N_{2RGBT}$), 3 RGBT star blends
($(N_{3RGBT}$), two or more RGBT star blends ($N_{\rm NRGBT}$), and
number of 2 RGB star blends ($(N_{2RGB}$) expected on the entire frame.
The last column is the total $K-$band magnitude sampled by one
resolution element.

\begin{deluxetable}{cc|cc|cccc|c}
\tablewidth{14cm}
\tablecaption{Theoretical Population of a NIC2 M31 Field}
\tabletypesize{\footnotesize}
\tablehead{
\colhead{$\mu_K$}			&
\colhead{L$_T$}				&
\multicolumn{2}{c}{\underline{$N_j$ / Res. Element}}&
\multicolumn{4}{c}{\underline{\hspace*{1.7cm} $N_{blends}$ / Frame \hspace*{1.7cm}}}&
\colhead{$M_K$}				\\
\colhead{(mag arcsec$^{-2}$)} &
\colhead{(\lsun)}	&
\colhead{RGBT}		&
\colhead{RGB}		&
\colhead{2RGBT}		&
\colhead{3RGBT}		&
\colhead{NRGBT}		&
\colhead{2RGB}		&
\colhead{}		}
\startdata
13 & 5340 & 0.6    & 70    & 7334      & 4308      & 17766      & \nodata & -7.02 \\
14 & 2126 & 0.2    & 28    & 1162      &  272      &  1517      & \nodata & -6.02 \\
15 &  846 & 0.09   & 11    &  184      &   17      &   203      & \nodata & -5.02 \\
16 &  337 & 0.04   &  4    &   29      &    1      &    30      & \nodata & -4.02 \\
17 &  134 & 0.02   &  2    &    5      &    0.07   &     5      & \nodata & -3.02 \\
18 &   53 & 0.006  &  0.7  &    0.7    &    0.004  &     0.7    &   10561 & -2.02 \\
19 &   21 & 0.002  &  0.3  &    0.1    &    0.0003 &     0.1    &    1674 & -1.02 \\
20 &    8 & 0.0009 &  0.1  &    0.02   &    0.0000 &     0.02   &     265 & -0.02 \\
21 &    3 & 0.0004 &  0.04 &    0.003  &    0.0000 &     0.003  &      42 &  0.98 \\
22 &    1 & 0.0001 &  0.02 &    0.0005 &    0.0000 &     0.0005 &       7 &  1.98 \\
\enddata
\label{tab:theoretical_populations}
\end{deluxetable}

\begin{figure}
\epsscale{1}
\plotone{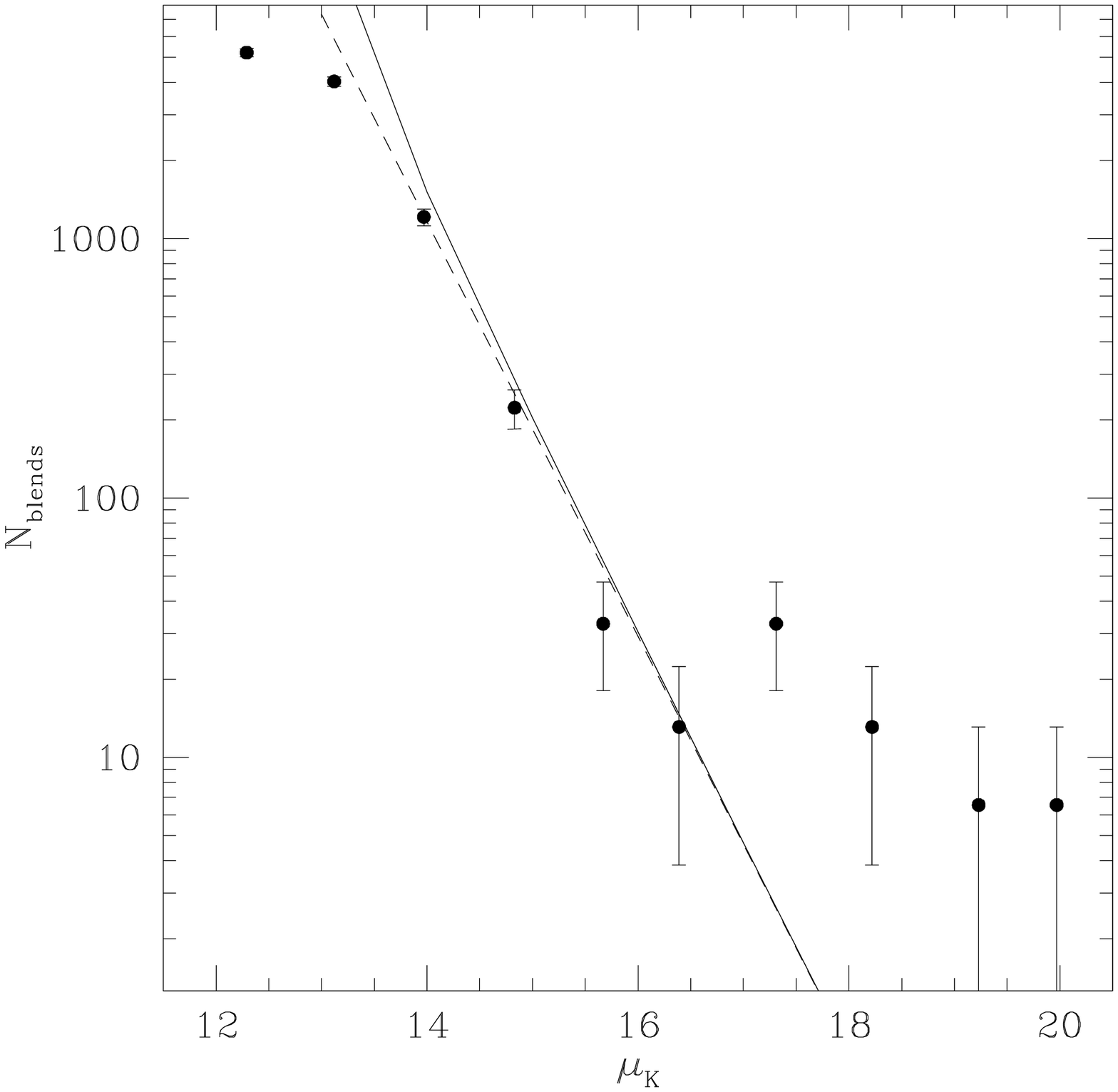}
\figcaption{
The solid line is the number of predicted blends on a NIC2 frame as a
function of surface brightness.  The dashed line shows the expected
number of only 2-star RGBT blends.  The points are the number of objects
measured on each mini-field which are brighter than the brightest input
star, normalized to the size of the NIC2 field.
\label{fig:blends}}
\end{figure}

The result is graphically shown in Fig. \ref{fig:blends}.  This plot
shows the predicted number of blends of 2 RGBT stars ($N_{2RGBT}$ --
dashed line), and 2 or more RGBT stars ($N_{NRGBT}$ -- solid line) on
the entire NIC2 frame.  Over plotted are the number of stars measured
brighter than the maximum input luminosity ($L_{\rm max}$) in the
simulations described in Section \ref{sec:artificial_fields}.

There is remarkable agreement for $14<\mu_K<17$.  Brighter than this,
the simulations are reaching the limit of the maximum number of stars
which can be fit on a frame, and hence fall they short of the
theoretical prediction.  For fainter surface brightnesses the flattening
in the simulations is due to a few objects which are recovered only
slightly brighter than $L_{\rm max}$, being blends of an RGBT star with
a much fainter star.

The general (qualitative) criterion for safe stellar photometry in
crowded field states that ``reliable photometry can be obtained only for
those stars that are brighter than the average luminosity sampled by
each resolution element" \citep{Ren1998}.  The last column in Table
\ref{tab:theoretical_populations} gives the total $K-$band magnitude
sampled in M31 by our resolution element. A comparison with
Fig. \ref{fig:artfieldsb} shows indeed that the criterion is well
confirmed by the detailed simulations.  This fairly obvious criterion
has been often violated in the past, resulting in a systematic
overestimate in the number (if any) of stars brighter than the RGB tip
in various stellar systems.

%
%
\section{Implications for Future Space-Based Observations} \label{sec:implications}

Even with the improved resolution of new space-based observatories,
measurements of distant objects will still be subject to the limitations
of image blending.  Since the blending threshold is determined by the
number of stars present in a resolution element, we can scale our
results according to the telescope PSF size and target distance to
obtain estimates of the limiting surface brightness for different
observations.  We need only assume that we have determined the limiting
number of stars per resolution element with our NICMOS simulations, and
that this limit will be similar for other space-based observatories,
i.e. assuming similar noise characteristics, PSF structure, and
performance as NICMOS.
\footnote{These results are of course also dependent upon having similar 
luminosity functions, and hence similar age.}

As an example, consider measuring Cepheid variables in external
galaxies.  Cepheids are one of the most important tools in determining
extragalactic distances; accurate determinations of their periods and
apparent magnitudes are one of the most important techniques for
obtaining distances to nearby galaxies.  Assuming a mean $K$-band
Cepheid luminosity of $M_K \sim -6$, our simulations show that NICMOS
observations of such stars at the distance of M31, the $K$-band surface
brightness must be fainter than $\mu_K \sim 15$ mag/arcsec$^2$ to make
accurate measurements.  This corresponds to a galactocentric distance of
$2'$ using the $r$-band \citet{Ken1987} surface brightness measurements,
and assuming $(r-K) \sim 2.8$ for a bulge population \citep{TDFD1994}.

What about Cepheid observations in the Virgo cluster?  Scaling our M31
$K$-band NICMOS results to the 15.9 Mpc distance of Virgo, we find that
the brightest background where NICMOS can accurately measure Cepheids is
$\mu_K=21.3$.  One example of a Virgo spiral is NGC 4548, which in fact
has been imaged by the HST Key Project team with WFPC2 \citep{Gra1999}.
The $V$-band surface brightness measurements of \citet{Ben1976},
indicate that the Key Project observations occurred in regions with
surface brightness $\mu_V \sim 22$.  Assuming $(V-K) \sim 1.5$ for a
young disk population \citep{PACF1983}, colors typical of SWB III
clusters in the LMC in which Cepheids are found, these regions have a
$K$-band surface brightness of $\mu_K \sim 20.5$ mag/arcsec$^2$.  Thus
the regions imaged with WFPC2 may be too dense for accurate NICMOS
photometry due to the infrared instrument's larger PSF.

Now consider future observations such as may be performed with a
space-based, 8-meter class telescope like the NGST.  The observations of
Cepheids in the Virgo cluster, which are difficult for NICMOS, will now
be trivial.  Again scaling our M31 NICMOS observations to the
anticipated NGST $H$-band PSF which will have a FWHM $\sim 0 \farcs 04$,
we find that the limiting $K$-band surface brightness for accurately
measuring Cepheid variables is $\mu_K \sim 18.4$.  In NGC 4548 this
corresponds to a distance of only $\sim 10''$ from the nucleus.

More distant targets, such as galaxies in the Coma cluster, will of
course be more challenging.  To scale from NICMOS observations in M31 to
NGST in Coma, we again assume that the $H$-band NGST PSF has a FWHM
$\sim 0 \farcs 04$, and that the Coma cluster is at a distance of
102Mpc.  This yields a limiting surface brightness of $\mu_K \sim 22.4$
for accurately measuring Cepheids.  Extrapolating the SB analysis of
\citet{Ken1987}, this limiting SB occurs at $\sim 100'$ from the center
of M31 (assuming $(r-K)=2.0$).  Thus placing M31 at the distance of the
Coma cluster, we find that the inner $0.7'$, or $\sim 26$kpc (assuming
$1'= 250$pc in M31), will be inaccessible to even NGST for accurately
measuring Cepheids in the infrared.

More detailed calculations should of course be undertaken when planning
observations, but these scaling arguments give a rough idea of what to
expect in terms of the effects of blending.

\begin{deluxetable}{ccccc}
\tablewidth{10cm}
\tablecaption{Cepheid Blending Limits}
\tabletypesize{\footnotesize}
\tablehead{
\colhead{Target}	&
\colhead{Distance}	&
\colhead{Telescope}	&
\colhead{FWHM}		&
\colhead{$\mu_K$(lim)} \\
\colhead{}		&
\colhead{(Mpc)}		&
\colhead{}		&
\colhead{($''$)}	&
\colhead{(mag/arcsec$^2$)}
}
\startdata
M31   &   0.7  & HST-NICMOS ($K$)  & 0.19  & 15.0  \\
M31   &   0.7  & NGST              & 0.04  & 11.6  \\
Virgo &  15.9  & HST-NICMOS ($H$)  & 0.15  & 21.3  \\
Virgo &  15.9  & NGST              & 0.04  & 18.4  \\
Coma  & 102    & NGST              & 0.04  & 22.4  
\enddata
\label{tab:cepheid_blend_limits}
\end{deluxetable}

%
%
\section{Summary \& Conclusions} \label{sec:conclusions}

In order to interpret out HST-NICMOS observations of globular clusters
in M31, we have developed techniques for understanding and quantifying
the effects of blending on stellar photometry.

Traditional completeness tests, specifically the injection and recovery
of a handful of artificial stars into a preexisting frame, are fine for
estimating the photometric completeness.  Unfortunately, this type of
test does not help in understanding the true brightnesses of stars seen
in crowded regions.  In very crowded regions, the few added artificial
stars are merely added on top of preexisting clumps, with no way of
recovering the composition of those clumps.

Our solution is to simulate the entire frame.  Our cluster simulations,
consisting of hundreds of thousands of stars each (Fig. \ref{fig:a280}),
revealed that stars as bright as we observe near the cluster centers,
can be easily created by the blending of many fainter stars
(Fig. \ref{fig:a280cmd}).  These simulations can be tailored to nearly
any observation, and use any desired input LF to best determine the true
properties of the observed stellar population, although where blending
is severe, there are degeneracies in the shape of the input LF.

Using uniformly populated fields of varying surface brightness, we
measured the degradation of photometric accuracy as a function of SB.
We determined that the photometry of faint stars ($M_K \sim -3$) began
to be affected by blending at a surface brightness of $\mu_K \sim 16$
magnitudes arcsecond$^{-2}$.  At a surface brightness of $\mu_K \sim 14$
the photometry of even the brightest stars was affected by blending
(Fig. \ref{fig:artfieldsb}).  With these criteria, we determined
threshold- and critical-blending radii for each cluster, which determine
the proximity to each cluster where reliable photometry can be achieved
(Tab. \ref{tab:blend_radii}).

We show the effects of blending on the measured luminosity function.  At
low surface brightnesses the measured LF accurately reflects the input
LF.  However with increasing SB, the measured LF becomes skewed toward
the bright end.  At very high SBs blending dominates the photometry, and
the measured LF has a slope opposite in sign to the input LF, actually
increasing toward bright stars (Fig. \ref{fig:artfieldlfs}).

Blending also affects the measured slope and width of the giant branch.
We quantified these effects, and determined a relation between the field
SB and the artificial increase in the metallicity as derived from the
RGB slope (Fig. \ref{fig:gbslope}).  Results of this modeling will be
applied to the cluster and field observations in Paper II.

By scaling our limiting surface brightness imposed by blending, we
estimate the limitations blending places on future space-based
observations.  We show that infrared observations of Cepheid variables
with NGST will be trivial in the Virgo cluster, but may be nearly
impossible in the Coma cluster.

\

\acknowledgements
Support for this work was provided by NASA through grant number GO-7826
from the Space Telescope Science Institute.  RMR acknowledges additional
support from NASA contract NAG5-9431, awarded in connection with the
NGST ad hoc science working group.  We thank Peter Stetson for supplying
and helping us with his photometry package, ALLFRAME.  Valuable comments
from Darren DePoy, Paul Martini and Marcia Rieke were greatly
appreciated.


\end{document}